\documentclass{aastex}

\slugcomment{Submitted to ApJ, 12 August 2009: revised November 2009}

\shorttitle{Variability in PPNs: I. C-Rich}

\shortauthors{Hrivnak et al.}

\begin{document}


\title{VARIABILITY IN PROTO-PLANETARY NEBULAE: I. LIGHT CURVE STUDIES 
OF 12 CARBON-RICH OBJECTS}


\author{Bruce J. Hrivnak, Wenxian Lu, Richard E. Maupin\altaffilmark{1}, and Bradley D. Spitzbart\altaffilmark{2}} 
\affil{Department of Physics and Astronomy, Valparaiso University,
Valparaiso, IN 46383; bruce.hrivnak@valpo.edu. wen.lu@valpo.edu}

\altaffiltext{1}{Present address: Department of Mathematics and Computer Science, University of Indianapolis,
Indianapolis, IN 46227, USA; rmaupin@uindy.edu}
\altaffiltext{2}{Present address: Smithsonian Astrophysical Observatory, High Energy Astrophysics,
60 Garden Street, MS-34, Cambridge, MA 02138, USA; bspitzbart@cfa.harvard.edu}

\begin{abstract}

We have carried out long-term (14 years) V and R photometric monitoring of 12 carbon-rich proto-planetary nebulae.  The light and color curves display variability in all of them.
The light curves are complex and suggest multiple periods, changing periods, and/or changing amplitudes, which are attributed to pulsation.
A dominant period has been determined for each and found to be in the range of 
$\sim$150 d for the coolest (G8) to 35$-$40 d for the warmest (F3).
A clear, linear inverse relationship has been found in the sample between the
pulsation period and the effective temperature and also an inverse linear relationship
between the amplitude of light variation and the effective temperature.
These are consistent with the expectation for a pulsating post-AGB star evolving
toward higher temperature at constant luminosity.
The published spectral energy distributions and mid-infrared images show these 
objects to have cool (200 K), detached dust shells and published models imply
that intensive mass loss ended a few thousand years ago. 
The detection of periods as long as 150 d in these requires a revision in the 
published post-AGB evolution models that couple the pulsation period to the 
mass loss rate and that assume that intensive mass loss ended when the pulsation
period had decreased to 100 d.  This revision will have the effect of extending
the time scale for the early phases of post-AGB evolution.
It appears that real time evolution in the pulsation periods of individual objects  may be detectable on the time scale of two decades.

\end{abstract}


\section{INTRODUCTION}

Proto-planetary nebulae (PPNs) are objects in transition between
the asymptotic giant branch (AGB) and planetary nebula (PN) phases
of stellar evolution. 
In this phase, the high AGB mass loss has ended and the star is 
surrounded by a detached, expanding shell of gas and dust.
Prior to the {\it Infrared Astronomical Satellite} ({\it IRAS}) satellite in 1983,
this phase remained as an observational ``missing link'' in the
post-main sequence evolution of stars of intermediate (1$-$8
M$_\sun$) mass. However, the use of the {\it IRAS} database has
allowed the identification of PPNs on the basis of
their infrared excesses and the colors of their circumstellar dust
shells \citep{partha86, veen89, hri89}. This has led to the
identification of $\sim$50 PPNs and an equal number of additional
candidates \citep{hri94, meixner99, ueta00, garlar97}. 
They show a double-peaked spectral energy distribution (SED), 
with a peak in the mid-infrared from re-radiated emission from the 
circumstellar dust and a peak in the visible$-$near-infrared from the 
reddened photosphere.
The spectral types range from B0 to K0.
Reviews of the properties of PPNs have been published by
\citet{kwok00} and \citet{hri03} and they are included in the 
more general review of post-AGB stars by \citet{vanwin03}. 

Millimeter radio observations of CO, OH, and HCN have reveled
the oxygen-rich or carbon-rich chemistry of the circumstellar 
shells \citep{lik89,lik91,omont93}.
High-resolution visible spectroscopy has 
shown the objects to be iron-poor and led to the determination of
the chemistry of the photospheres \citep{vanwin03}.  
High-resolution imaging with the {\it Hubble Space Telescope} ({\it HST}) 
has revealed the morphology of the
nebulae, which are generally elliptical or bipolar with sizes
up to a few arcseconds in radius \citep{ueta00,su01,sahai07,siod08}.

We are carrying out a long-term study of the variability of PPNs.
It initially began with radial velocity monitoring to search for 
possible binary companions.  This was motivated in part by the 
idea that a binary companion could cause the dusty torus 
seen in bipolar PPNs, which apparently restricts the outflow and
results in the bipolar morphology \citep{livio88}.
The results of this initial study did not reveal any binaries, but
did reveal periodic velocity variations \citep{hri00a,hri09e}.

The radial velocity monitoring was followed-up by the initiation 
of a study of the light variations in PPNs, which has been ongoing 
for the past 15 years.  
We identified an initial sample of $\sim$40 PPN candidates and have
monitored them for light variability.  
All of the PPNs are found to be light variables, and 
the preliminary results have been presented \citep{hri00a,hri02,hri07a}.
Arkhipova and collaborators \citep[e.g.,][]{ark00} have also presented the results of 
their light curve studies of several PPNs .

The results of our ongoing studies will be published in a series of
papers.  In this initial publication, we focus on 12 carbon-rich PPNs.
We present in detail our observing program and discuss the 
observed light curves, which we investigated for periodicity.  
The results are then discussed in the broad context of post-AGB
variability and evolution.
In two companion papers, radial velocity data are presented for 
four of the brightest of these 12 PPNs, and together they are used in the 
interpretation of the discovered pulsational variability \citep{hri09a, hri09b}. 

\section{PROGRAM OBJECTS}

The program objects are listed in Table~\ref{object_list}. They
are all relatively bright, V = 8 to 14 mag, and thus are
accessible to study with a small telescope.  
All of the targets with the exception of IRAS 19500$-$1709 have been 
imaged with the {\it HST} and all but one show elongated nebulae 
and faint halos with the central stars visible  \citep{ueta00,sahai07,siod08}.  
Only for AFGL 2688 is the central star not visible; this object is seen 
in visible light only by scattering from its bipolar lobes and halo.  
About half of the targets have published mid-infrared images taken
with 6$-$10 m telescopes \citep{kwo02,ueta01,clube04,mor00,clube06}.

\placetable{object_list} 

All 12 of the objects appear to be bona-fide PPNs. 
They all have double-peaked SEDs  
and their spectra are classified as F$-$G supergiants.  Their carbon-rich
nature was first recognized by the detection of C$_2$ and C$_3$
absorption in their visible spectra \citep{hri95,bak97}. This
has been confirmed with higher resolution abundance studies,
which find an average value of C/O = 1.6 \citep{vanwin00,red02}.
These studies find the objects to be slightly metal poor, [Fe/H] = $-$0.7,
with high abundances of s-process elements such as Y, Ba, Ce, and Nd 
([s/Fe] = 1.5), which are the results of nucleosynthesis and the third-dredge 
up during the AGB phase. Infrared spectra have shown these objects to
display the aromatic infrared bands (AIBs) in emission 
at 3.3, 6.2, 7.7, 8.6, and 11.3 $\mu$m and aliphatic emission bands 
at 3.4, 6.9 $\mu$m \citep{hri00,hri07}, 
commonly attributed to PAH molecules.  Also seen in their
mid-infrared spectra are a broad emission feature at 20.1 $\mu$m
\citep[the ``21 $\mu$m'' emission feature;][]{vol99,hri09} and a very broad
emission feature around 30 $\mu$m \citep[the ``30 $\mu$m'' emission
feature;][]{vol02,hon02}, which are found only in carbon-rich evolved
objects, particularly PPNs. These spectroscopic properties are summarized in
Table~\ref{object_prop}.  (See \citet{hri08a} for a more detailed table of
properties.)  

\placetable{object_prop}

Our object list composes an almost complete sample of known carbon-rich PPNs
north of decl. $-$20$\arcdeg$ and brighter than V = 14.5 mag.  The only 
exceptions are IRAS 06530$-$0213 (F5I, V=14.2), which was too faint 
for this study at its southerly declination, and IRAS 01005+7910 
(B0I, V=10.9), which varies with a very short time scale and which
will be discussed in a future study.

\section{OBSERVATIONS AND REDUCTIONS}

Photometric observations were made at the Valparaiso University
Observatory (VUO).  The observatory is equipped with an 0.4-m,
f/12 computer-controlled telescope made by DFM Engineering, Inc.
The detector was a Photometrics star I 576$\times$384 CCD with a uv-enhanced
coating and a field of view of 8$\farcm$6$\times$5$\farcm$7.  
A filter set was used that provided a good match to the standard BV (Johnson) and RI
(Cousins) systems.

Some preliminary observations of several of the sources were made
in the summer of 1993 which indicated their variability. The
observing program was begun in earnest in the summer of 1994, with
observations made using only the V filter.
Beginning in the summer of 1995, occasional observations were made
through the R filter to investigate color variations, and these
were continued with increased frequency.  Since summer 2002,
the V and R observations have consistently been made together. The
typical annual observing schedule involved frequent observations
of the objects available in the summer and fall and infrequent
observations during the winter and spring.  Unfortunately a defect
developed in the V filter in 2000 that remained undetected for a
long time.  This has led us to reject all of the V data from
2451600 to 2452508 (2000 Feb 25 day to 2002 Aug 21), 
resulting in a gap in the V light curves during that interval.

Integration times were adjusted with a goal of obtaining a
precision of  $\sigma$ $\le$ $\pm$0.01 mag.  This precision was
obtained for all but the faintest sources; integration times were
limited to a maximum of $\sim$6 min due to lack of
a guiding system for the telescope.

The observations were reduced using standard procedures in
IRAF\footnote{IRAF is distributed by the National Optical
Astronomical Observatory, operated by the Association for
Universities for Research in Astronomy, Inc., under contract with
the National Science Foundation.} to remove cosmic rays, subtract
the bias, and flat field the images. Aperture photometry was
carried out, using an aperture of 11$\arcsec$ diameter. This
accommodated the seeing quality of the site, which was typically
$\sim$3$\arcsec$. Only in the case of the fainter, southern lobe
of AFGL 2688 (``Egg'' nebula) was a smaller aperture
(7$\arcsec$) used; this was to avoid contamination by the brighter
northern lobe of this large, bipolar nebula.

Standardized photometry of several of these PPNs was carried out at the VUO on several
nights using stars from the list of \citet{landolt83}.  
We had previously obtained standardized photometry of most of them as part of 
our larger studies of PPNs and these were previously published.  
We have listed all of these photometric results in Table~\ref{std_ppn}.
Many have very red colors due to reddening by their circumstellar dust.

\placetable{std_ppn}

Since our goal was to investigate variability in these PPNs, the
observing program was organized around differential photometry of
the candidates with respect to non-variable comparison stars in
each field. This allowed us to collect data on the many
non-photometric nights available at our site. A set of three
comparison stars was established in each PPN field. 
The comparison stars were chosen in such a way as to 
minimize the standard errors in the differential magnitudes.
For the fainter PPNs, they ranged from a little to a lot brighter 
and for the brighter PPNs they were a lot fainter than the PPNs;
for the bright PPNs we were more limited in our choices.
Given the redness of our highly reddened PPNs compared to
the comparison stars, the color matches of the PPNs and the
comparison stars were usually not very close except for 
the F spectral type PPNs.
The constancy of the primary comparison star C1 in each field 
was established over this interval. In a few cases, small variability 
in C2 or C3 is indicated. 
The level of constancy of the comparison stars C1 with respect to C2 
is displayed in the subsequent light curve figures. 
The identity of the comparison stars and their standardized photometric
magnitudes (when available) are listed in Table~\ref{std_comp}.

\placetable{std_comp} 

The differential V, R, and (V$-$R) 
measurements of each PPN with respect to its primary comparison star
C1 were transformed to the standard magnitude system and 
are listed in Tables~\ref{tab_22223} through 17. 
Although in many cases we do not have a simultaneous set of V and 
R observations, we were able to carry out the standardization as follows.
The color coefficients for standardization are small, +0.030 for V and 
+0.049 for R, and thus our instrumental differential magnitudes 
are close to the standard system.  
Since none of the program stars vary much in color,
$\Delta$(V$-$R) $\le$ 0.12 mag  (except 0.18 mag for IRAS 05113+1347), 
a simple zero-point correction will bring the differential
magnitudes close to the standard system.  
However, since there is a correlation between brightness and color
(see Section~\ref{lc_trends}), we were able to use this to make a very good 
approximation to correct for the star's color at each observation.
Differential extinction was
small and was not applied to the observations.  
Uncertainties were
calculated for the differential measurements and are shown in the
light curve plots to follow.  We have not listed them in the
tables, but to give an idea of the precision of the observations, 
we have tabulated the maximum and average uncertainties for each 
PPN in Table~\ref{tab_stat}, along with the number of observations.

\placetable{tab_22223}

\placetable{tab_stat}

\section{RESULTS}

\subsection{Discussion of Individual Objects}

The objects are discussed individually in the following text.
We begin with the clearest cases of periodic variability and proceed 
in order of right ascension, beginning with IRAS 22223+4327.

\begin{figure}\figurenum{1}\epsscale{0.9}
\plotfiddle{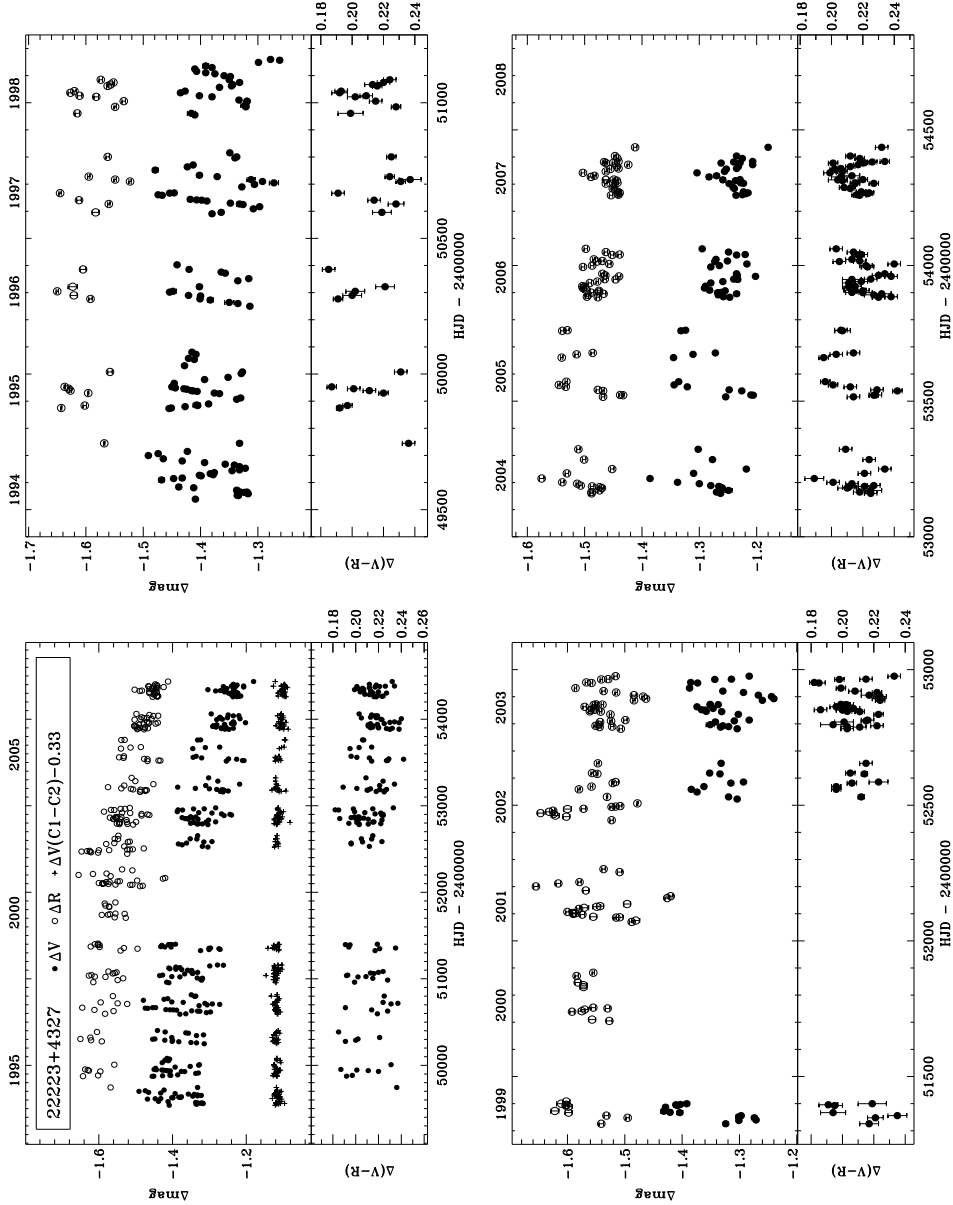}{0.0in}{+180}{400}{500}{0}{0}
\caption{Plot showing the differential light and color curves of IRAS 22223+4327.  In the upper left panel are the entire curves and in the three other panels the curves are shown on expanded scales covering five years each.  The error bars are included with the data in these expanded plots.  
Also shown in the upper left panel is the differential V light curve of the primary (C1) with respect to the secondary (C2) comparison stars on the same scale as the PPN light curves. 
Zero-point offsets are added to conveniently show all three light curves on the same plot. 
\label{22223_lc}}\epsscale{1.0}
\end{figure}

{\it IRAS 22223+4327} (Figure~\ref{22223_lc}).$-$
This object displays a light variation with a clear cyclical
pattern of moderate amplitude during each year, 
superimposed on a general dimming of the object. 
The individual differential observations are of good
precision ($\sigma$(V)$\le$0.008 mag).  
The light curves show an overall monotonic decrease in brightness 
by $\sim$0.16 mag in 14 years.
Visual inspection
of the light curves suggests a period to the cyclical variation of $\sim$90 d. 
The full range in brightness varies from season to season, with a maximum
(peak-to-peak) in $\Delta$V of 0.21 mag (1997-98) and a minimum of
0.09 mag (2002-03, 2006-07).
An even greater brightness range is present in the 2000-01 and 2001-02 observations as
documented in $\Delta$R, from 0.065 (also in 2006-07) to 0.235 mag.
The color $\Delta$(V$-$R) varies over a range of 0.06 mag, 
and is systematically redder when the object is fainter
in its cyclical variation.  
A small seasonal change in color is seen over the past four seasons, 
with the system slightly redder (0.01-0.02 mag) when dimmer.

\placefigure{22223_lc}

In a companion paper \citep[Paper~II]{hri09a}, this periodic light curve is
analyzed together with the observed velocity curve to better understand the
pulsations in the object.  The light curves are discussed in more detail in that paper.

\begin{figure}\figurenum{2}\epsscale{0.9}
\plotfiddle{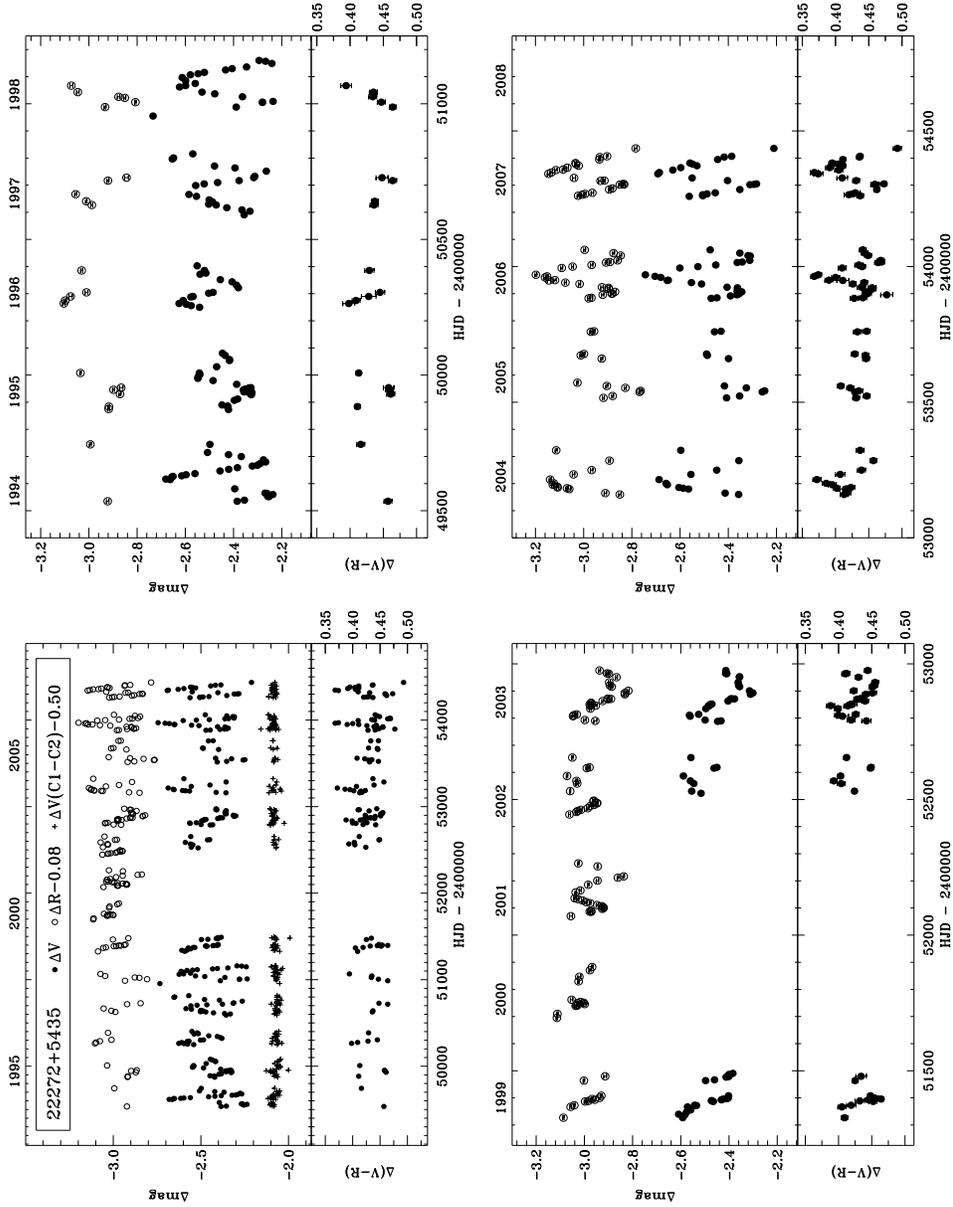}{0.0in}{+180}{400}{500}{0}{0}
\caption{Plot showing the differential light and color curves of IRAS 22272+5435, plotted similar to Figure~\ref{22223_lc}.  
\label{22272_lc}}\epsscale{1.0}
\end{figure}

{\it IRAS 22272+5435} (Figure~\ref{22272_lc}).$-$
This object displays a cyclical variation in brightness, but with
a large range in amplitudes, varying in $\Delta$V from a maximum 0.48-0.49 mag
in 1998-99 and 2007-08 to a minimum 0.22 mag in 1995-96, with a
full range of 0.53 mag. 
The individual differential observations are of good precision ($\sigma$(V)$\le$0.008 mag).
The observed range in R is not so great 
but there are fewer observations.  
The mid-range value changes slightly from season to season over a
range of $\sim$0.10 mag in V.
A period $\sim$130 d is suggested by visual inspection of the light curves. 
The seasonal variations are somewhat reminiscent of a beat period, 
suggesting a second period similar to the first.
The color varies systematically with brightness $-$ it is redder when fainter $-$
with a typical seasonal range of 0.08 mag and a maximum range of
0.12 mag.
The profile of IRAS 22272+5435 is clearly extended in our images, 
with a FWHM that is $\sim$10\% larger than the other stars in the field. 
This is consistent with the {\it HST} images, which show it to be extended with 
a large halo ($\sim$3.5$\arcsec$ in diameter).

\placefigure{22272_lc}

This object is also analyzed together with the observed radial
velocity curve in a companion paper with IRAS 22223+4327 
(Paper~II). 

\begin{figure}\figurenum{3}\epsscale{0.9}
\plotfiddle{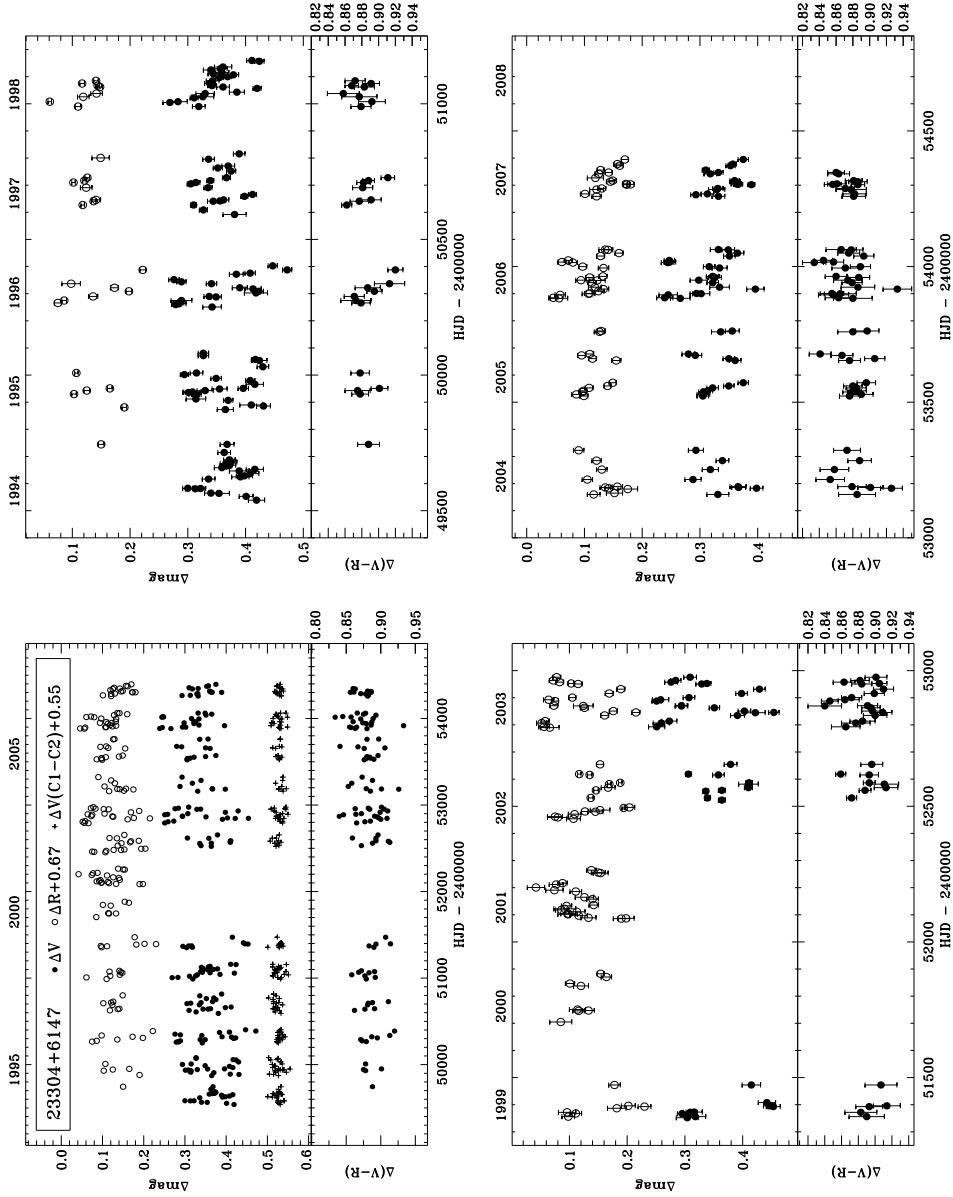}{0.0in}{+180}{400}{500}{0}{0}
\caption{Plot showing the differential light and color curves of IRAS 23304+6147, plotted similar to Figure~\ref{22223_lc}.  
\label{23304_lc}}\epsscale{1.0}
\end{figure}

{\it IRAS 23304+6147} (Figure~\ref{23304_lc}).$-$
This object displays a regular cyclical variation in light,
although in the first and fifth seasons it appears as if the
variation is damped out over approximately one cycle.
The object is relatively faint (V=13.1) and consequently the error bars
are relatively large (see Table~\ref{tab_stat}; typically $\pm$0.011 mag in V, 
$\pm$0.009 mag in R, $\pm$0.0013 mag  in (V$-$R)). 
The variation appears to have a period of $\sim$80 d and a
peak-to-peak maximum of 0.20 mag in V and 0.16 in R. 
The object is redder when fainter, but the typical seasonal color range 
is only 0.06 mag (maximum of 0.10 mag).  There appears to be perhaps a 
slight change in color over the 14 year observing interval, 
with the object becoming bluer by 0.02 mag in (V$-$R).

\placefigure{23304_lc}

\begin{figure}\figurenum{4}\epsscale{0.9}
\plotfiddle{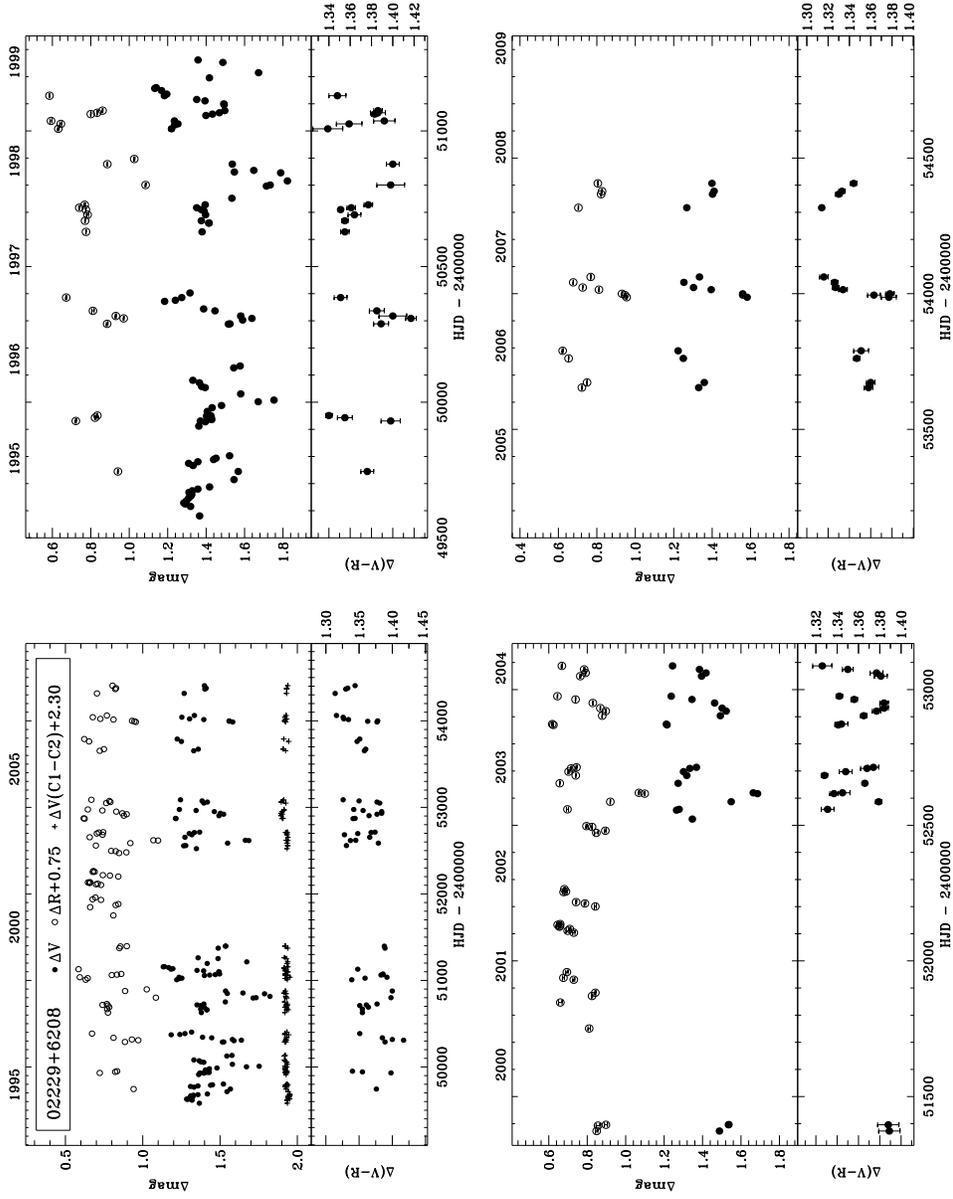}{0.0in}{+180}{400}{500}{0}{0}
\caption{Plot showing the differential light and color curves of IRAS Z02229+6208, plotted similar to Figure~\ref{22223_lc}.   \label{02229_lc}}\epsscale{1.0}
\end{figure}

{\it IRAS Z02229+6208\footnote{The letter Z is added at the beginning because the object 
is absent from the {\it IRAS} Point Source Catalog but is present in the Faint Source 
Reject File.  See \citet{hri99} for a discussion of the identification of this object. 
Note that we have omitted the {\it Z} in some of the figure labeling.}} 
(Figure~\ref{02229_lc}).$-$  
This object displays large, long-term variations 
of up to 0.54 mag (V) in a season.  
Measuring the time between successive 
minima results in cycle lengths of $\sim$130 to $\sim$160 days.  
The light curve varies in appearance from cycle to cycle, with a
suggestion that minima alternate between deeper and shallower ones,
but this does not appear to follow a strict pattern.  
The object is redder when fainter.  
The maximum color range in a season is 0.06 mag, but some of the
early seasons with large variations in V have few R observations
and the color is consequently poorly sampled.
The full range in V$-$R is 0.09 mag, and it appears that the
overall color of the system was getting slightly bluer (by 0.03 mag)
from 1995 to 2007.

\placefigure{02229_lc}

\begin{figure}\figurenum{5}\epsscale{0.9}
\plotfiddle{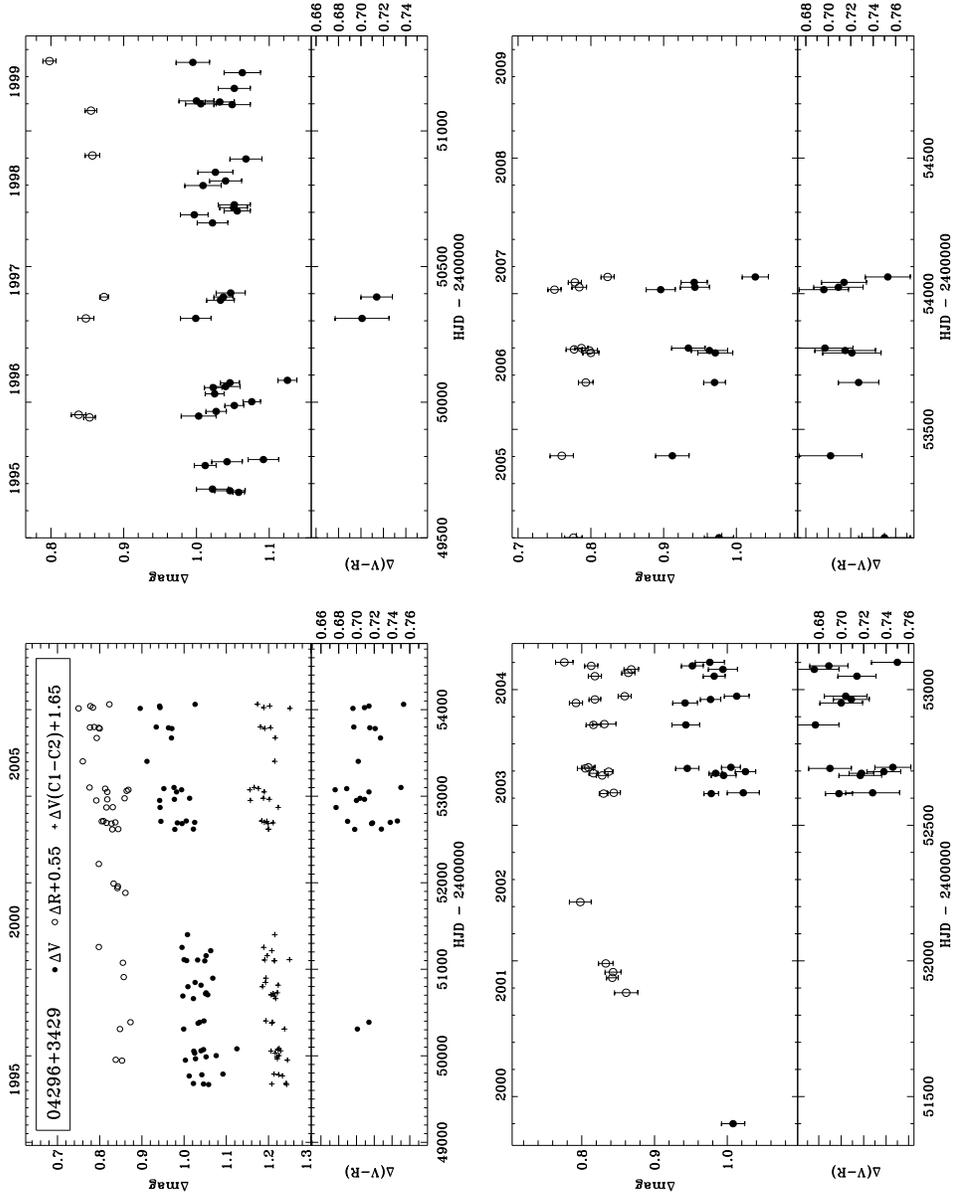}{0.0in}{+180}{400}{500}{0}{0}
\caption{Plot showing the differential light and color curves of IRAS 04296+3429, plotted similar to Figure~\ref{22223_lc}.   \label{04296_lc}}\epsscale{1.0}
\end{figure}

{\it IRAS 04296+3429} (Figure~\ref{04296_lc}).$-$
This object is one of the faintest targets in the present observing
 program ({\it V}=14.2), and the uncertainties are consequently large, 
 particularly in {\it V}.
 This is also seen in the scatter in the differential magnitudes of the
 comparison stars, since all of the stars in the field are relatively faint 
 ({\it V} $\ge$ 13.2 mag) for a 6 min exposure.
The system appears to show a monotonic increase in brightness of 
0.10 mag in V from 1994-95 to 2006-07.  
The increase is less in R, but the number of observations is smaller.  
The variation within a season ranges from 0.05 to 0.13 mag.
The large uncertainty compared to the range prevents us from 
saying much about the light curve variability or correlation with color.
However, in 2007, when the largest range in brightness is seen, 
the objects appears to be redder when fainter.
Note that we have begun to observe this object at higher
precision (with guiding) in a new observing program designed for fainter targets.

\placefigure{04296_lc}

\begin{figure}\figurenum{6}\epsscale{0.9}
\plotfiddle{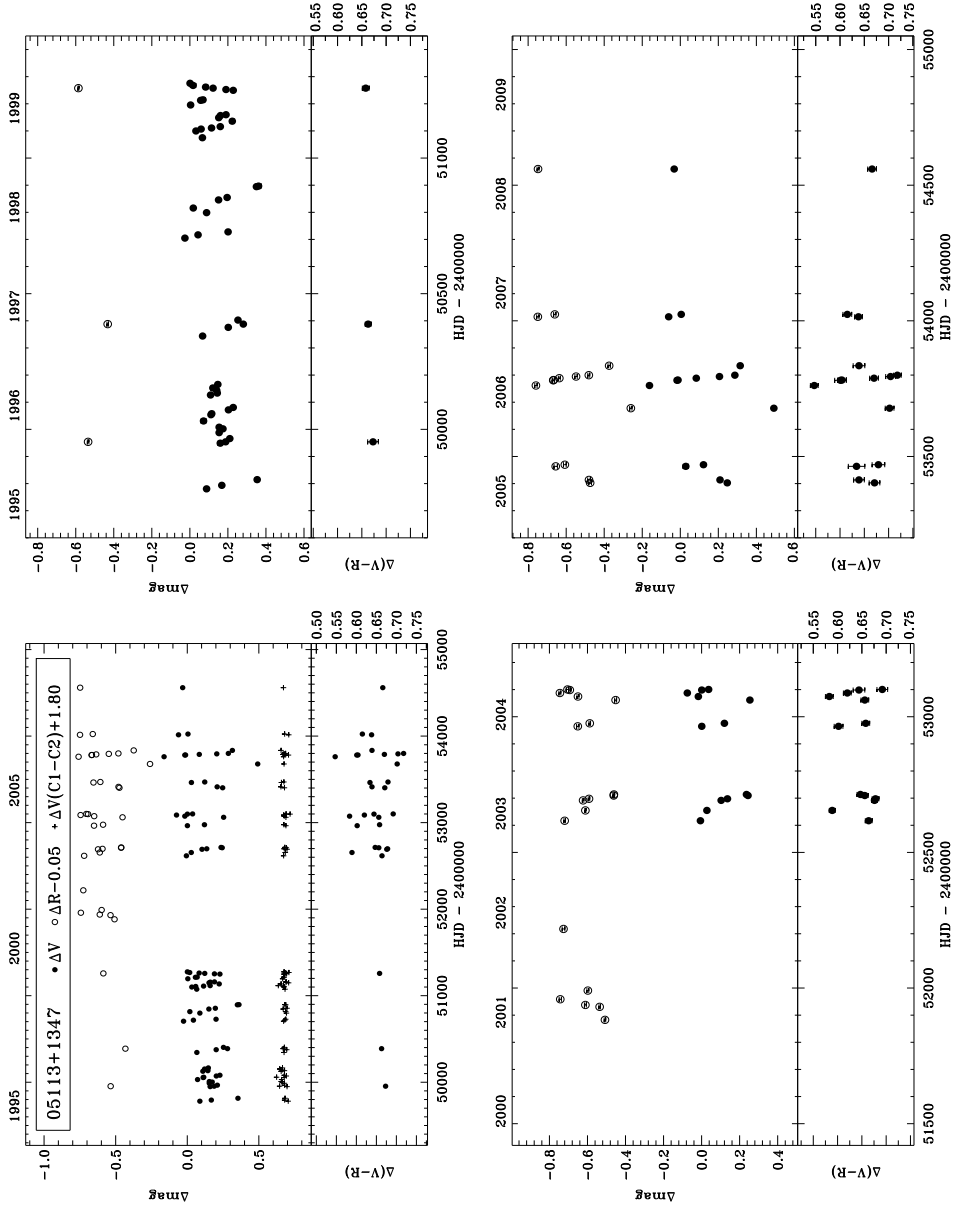}{0.0in}{+180}{400}{500}{0}{0}
\caption{Plot showing the differential light and color curves of IRAS 05113+1347, plotted similar to Figure~\ref{22223_lc}.   \label{05113_lc}}\epsscale{1.0}
\end{figure}

{\it IRAS 05113+1347} (Figure~\ref{05113_lc}).$-$
This object clearly varies, with 
peak-to-peak variations of 0.67 (V) and 0.49 (R) mag, 
seen in 2005-2006.  The variation appears to be cyclical, with cycle length 
of $\sim$120$-$130 d, 
and the amplitude of the variation clearly changes.  
The color changes over a range of 0.18 mag, seen also 
in 2005-2006, and the system is redder when fainter.  

\placefigure{05113_lc}

\begin{figure}\figurenum{7}\epsscale{0.9}
\plotfiddle{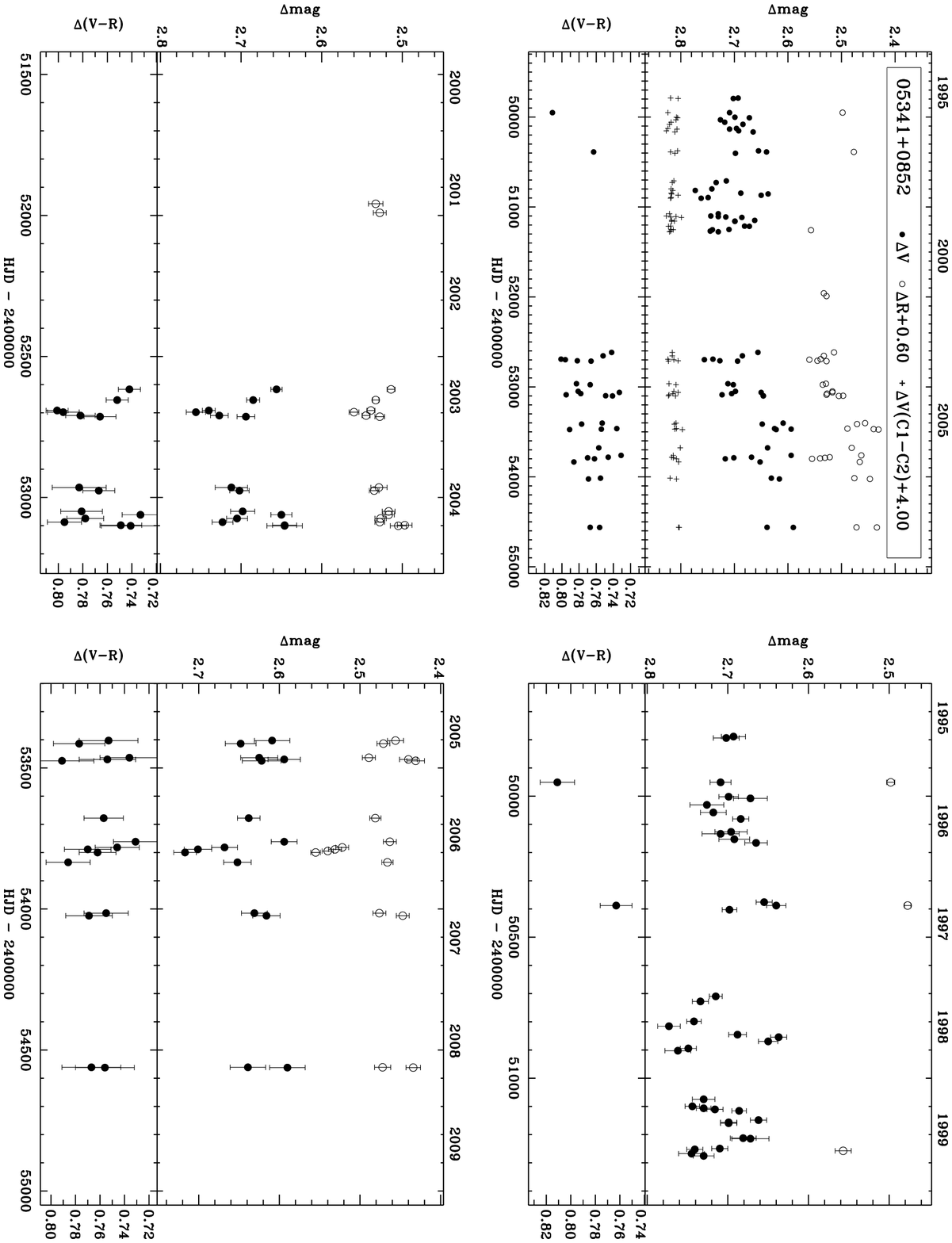}{0.0in}{+180}{400}{500}{0}{0}
\caption{Plot showing the differential light and color curves of IRAS 05341+0852, plotted similar to Figure~\ref{22223_lc}.   \label{05341_lc}}\epsscale{1.0}
\end{figure}

{\it IRAS 05341+0852} (Figure~\ref{05341_lc}).$-$
The object is faint (V=13.6 mag), and the uncertainties are consequently large.
This object displays a peak-to-peak variation of 0.18 mag,
with the overall brightness appearing to increase by 0.09 mag beginning in 2005. 
A suggestion of a cyclical variation is seen in several seasons, with a maximum 
seasonal range of 0.13 mag (1997-98) and a suggestion of a period of $\sim$70 days in 
1997-98 and 1998-99.  
However, there are not a lot of observations.  
The color changes over a range of 0.08 mag and is generally redder when fainter, 
although the uncertainty in the data is large.

\placefigure{05341_lc}

\begin{figure}\figurenum{8}\epsscale{0.9}
\plotfiddle{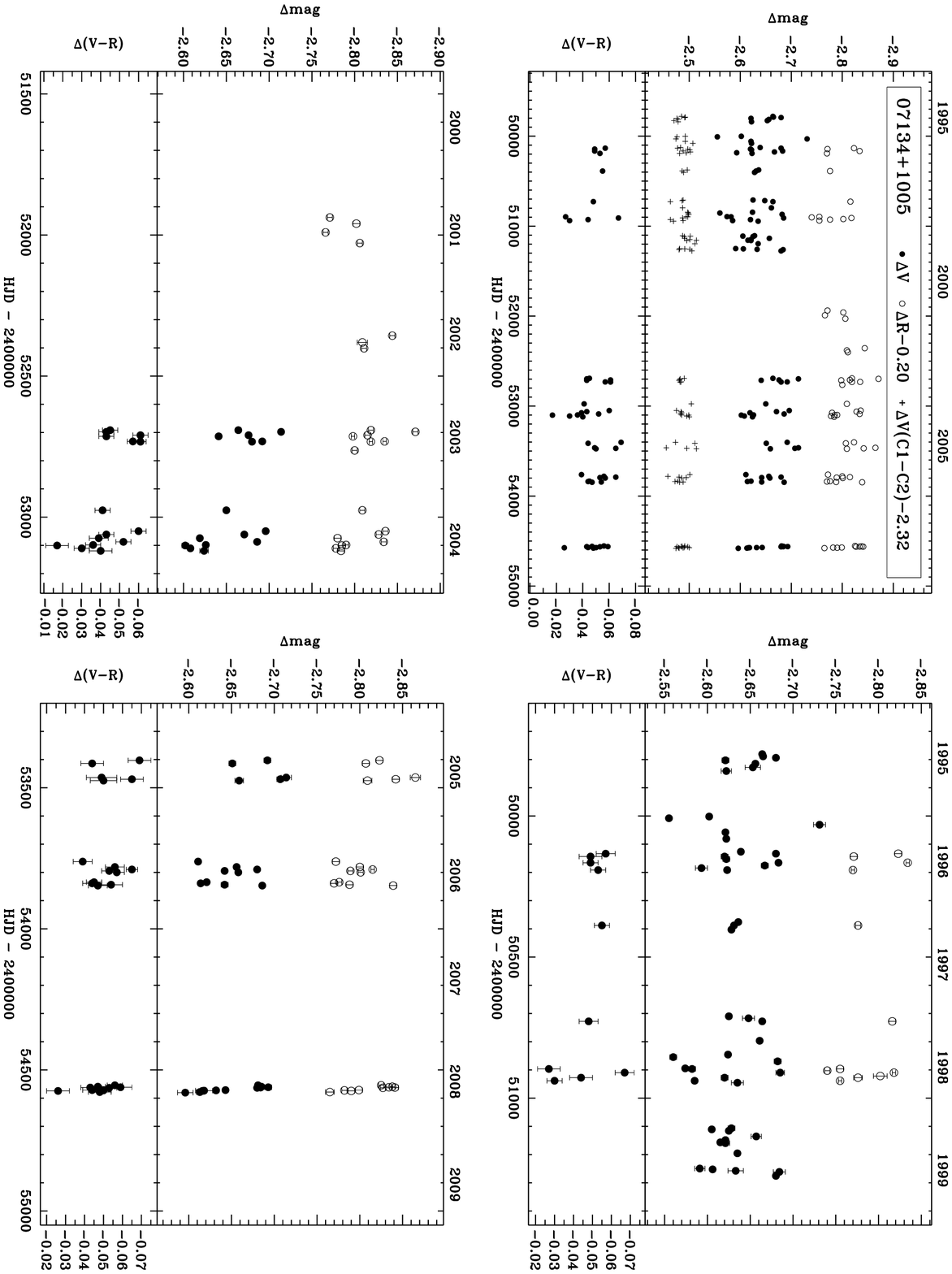}{0.0in}{+180}{400}{500}{0}{0}
\caption{Plot showing the differential light and color curves of IRAS 07134+1005, plotted similar to Figure~\ref{22223_lc}.   \label{07134_lc}}\epsscale{1.0}
\end{figure}

{\it IRAS 07134+1005} (Figure~\ref{07134_lc}).$-$  
This object clearly varies, with a typical 
seasonal range of 0.09 mag (V) and a maximum of twice that value in 1995-96.
The variations generally look cyclical, but no consistent period is obvious by
visual inspection.
The object is slightly redder when fainter, but with a peak-to-peak range
of only 0.045 mag.

\placefigure{07134_lc}

The light curve of this object is analyzed together with the observed radial
velocity curve in a second companion paper \citep[Paper~III]{hri09b}.

\begin{figure}\figurenum{9}\epsscale{0.9}
\plotfiddle{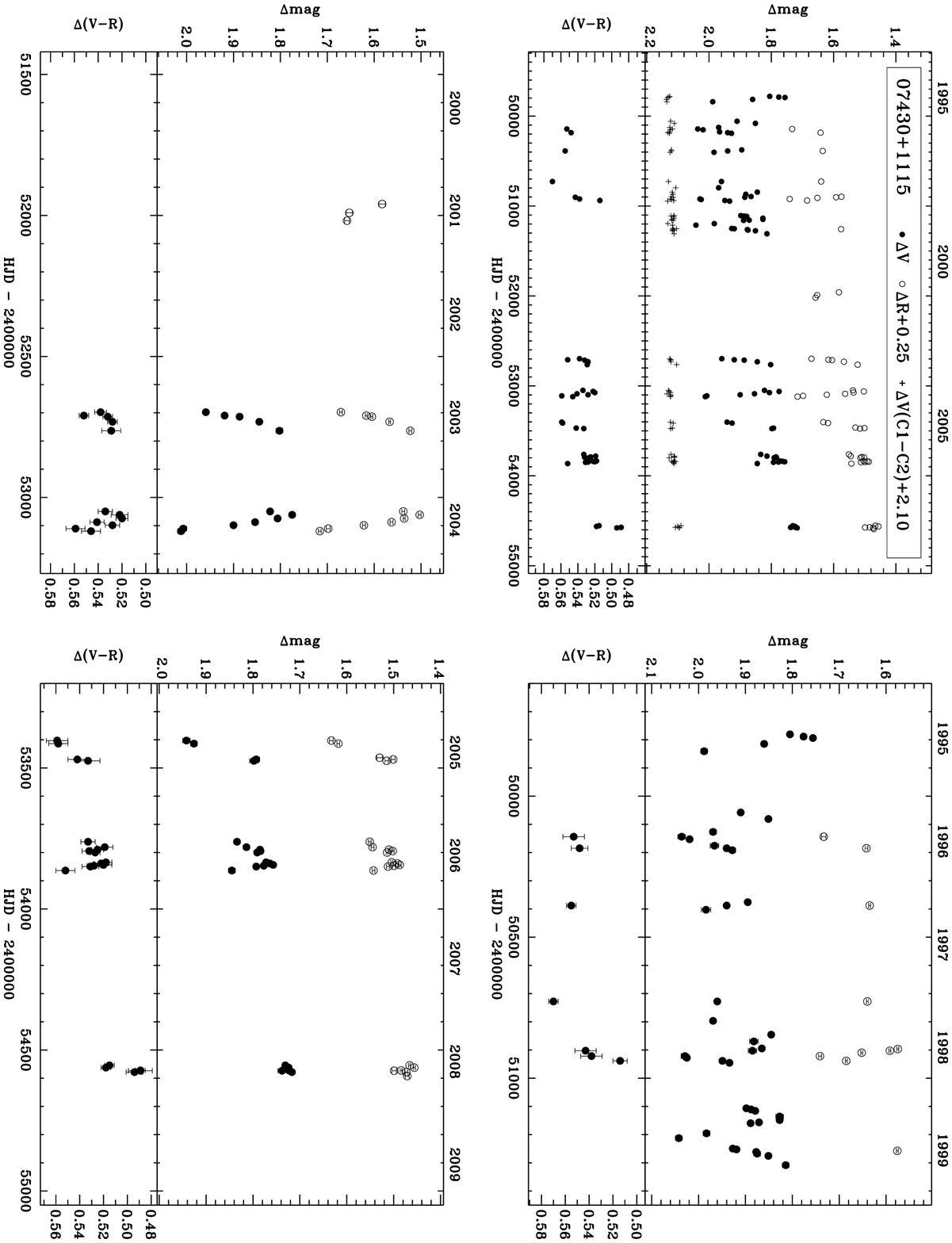}{0.0in}{+180}{400}{500}{0}{0}
\caption{Plot showing the differential light and color curves of IRAS 07430+1115, plotted similar to Figure~\ref{22223_lc}.   \label{07430_lc}}\epsscale{1.0}
\end{figure}

{\it IRAS 07430+1115} (Figure~\ref{07430_lc}).$-$
The object clearly varies, with ranges (peak-to-peak) of 0.32 mag in V 
and 0.27 in R, although part of the apparent difference in ranges may be due to the 
fewer R observations in the early years.  
The variations appear to be cyclical, with a typical seasonal range of 
0.15 to 0.23 mag (V) and a cycle of $\sim$150 d.  
The overall stellar brightness also appears to change, 
dimming by $\sim$0.1 mag between 1995 and 1996 and then gradually 
increasing to the initial 1995 value by $\sim$2004.
The color is redder when fainter, with a color range of 0.08 mag.
Some of the color change may correspond to the long-term seasonal change in
brightness and the rest to the cyclical variation, but these are hard to distinguish 
given the small number of color data points.

\placefigure{07430_lc}

\begin{figure}\figurenum{10}\epsscale{0.9}
\plotfiddle{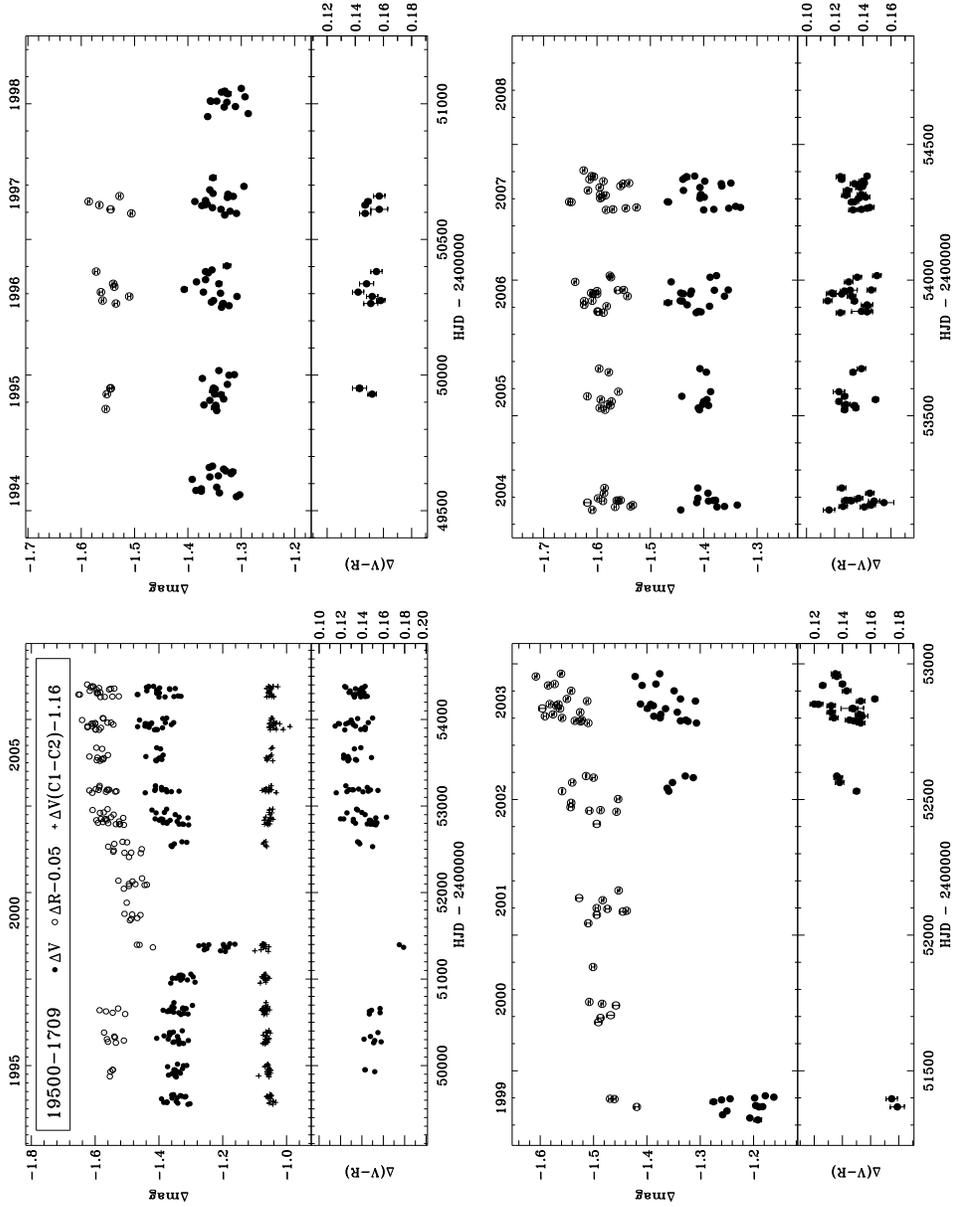}{0.0in}{+180}{400}{500}{0}{0}
\caption{Plot showing the differential light and color curves of IRAS 19500$-$1709, plotted similar to Figure~\ref{22223_lc}.   \label{19500_lc}}\epsscale{1.0}
\end{figure}

{\it IRAS 19500$-$1709} (Figure~\ref{19500_lc}).$-$
The star shows a cyclical seasonal variation with changing amplitude.  
In addition, the overall light level of the star has changed.
From 1993-1998, the mean brightness was approximately constant.
However, in 1999, it was found to be fainter by 
0.12 mag in V and 0.10 mag in R.  
It gradually brightened by about 0.19 mag in V and 0.14 mag in R
from then through 2004, and it has been at approximately 
the same mean brightness through 2008.
As is apparent by these brightness values, the color of the star 
also changed during this time, getting redder in (V$-$R) by
0.02 mag when the system dimmed and then getting bluer by 0.04 mag 
when it brightened.
(Note that photometric observations of this object by \citet{ark00} from
1993 to 1999 show a similar light and color curve change in 1999.)
The seasonal cyclical variations show amplitudes varying from 
0.05 to 0.13 mag in both V and R.
The maximum color range in a season is only 0.04 mag;
there is a slight tendency for the object to be slightly redder when fainter.
Visual inspection indicates cycle lengths of 30-35 days.

\placefigure{19500_lc}

This object is analyzed together with the observed radial
velocity curve in a companion paper with IRAS 07134+1005
(Paper~III).  

\begin{figure}\figurenum{11}\epsscale{0.9}
\plotfiddle{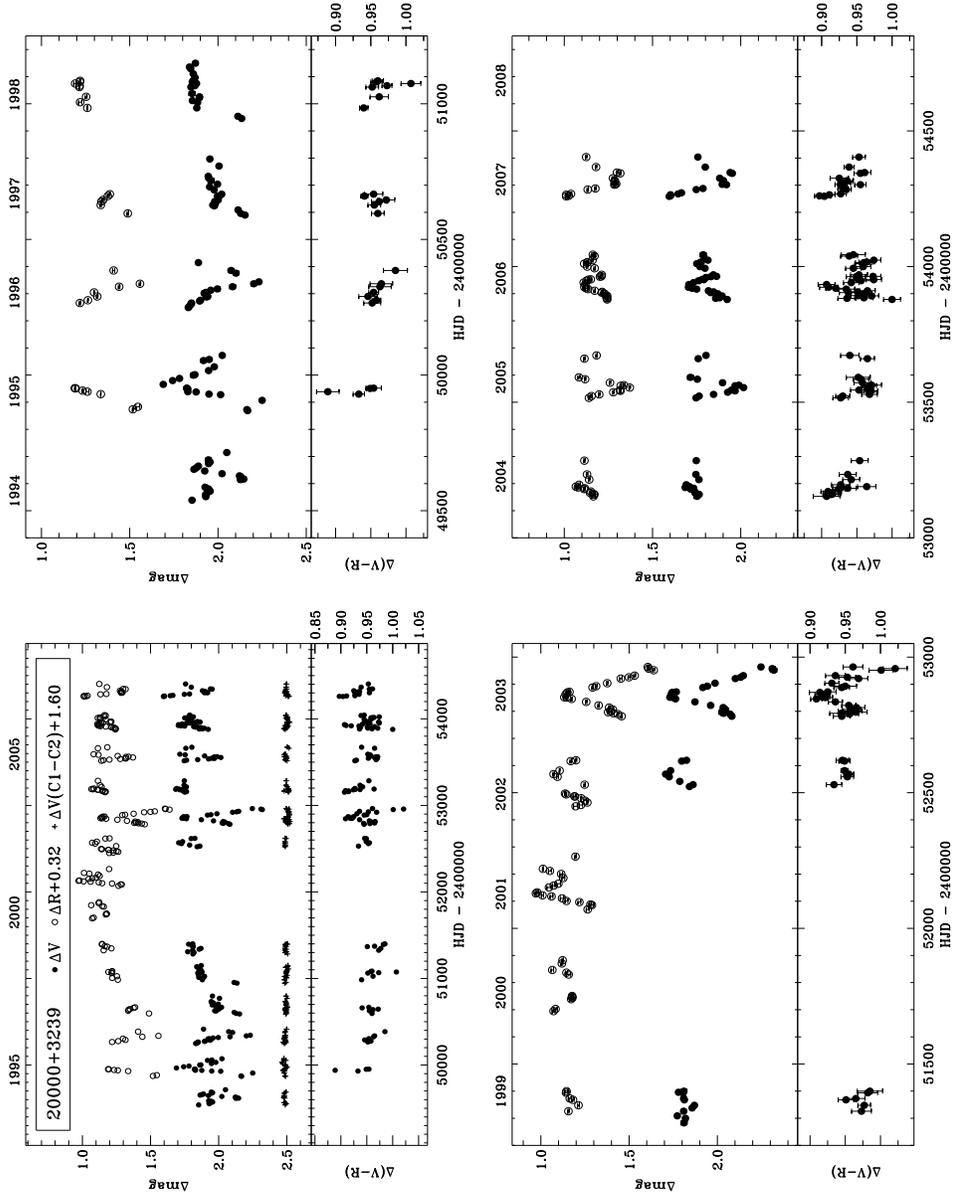}{0.0in}{+180}{400}{500}{0}{0}
\caption{Plot showing the differential light and color curves of IRAS 20000+3239, plotted similar to Figure~\ref{22223_lc}.  \label{20000_lc}}\epsscale{1.0}
\end{figure}

{\it IRAS 20000+3239} (Figure~\ref{20000_lc}).$-$ 
The light curve shows a large variation from season to season and within a season.
A cyclical variation is seen over the first three seasons (1994$-$1996) 
with deeper and shallower local minima,
somewhat similar to those seen in RV Tau variables. However, there
does not appear to be a distinct pattern of deeper and shallower minima.  
The variability range changes from year to year.
A cyclical variability is seen during the first three seasons, while
in the next two seasons
(1997$-$1998), the brightness appears to rise relatively quickly
to a plateau value.  
The amplitude of the cyclical variation then generally begins to increase to a 
maximum range in 2003 of 0.59 mag (V), but it varies quite a bit from 
year to year.  The maximum brightness level of the object is greater
after 1999 than before.
The variations in R are similar.
Color variations range over 0.12 mag, with a general trend of
being redder when fainter. 
The individual observations are of good precision, even though the object
is faint (V=13 mag), with typical values of $\sigma$(V) = 0.010 mag, 
$\sigma$(R) = 0.005 mag, and $\sigma$(V$-$R) = 0.011.

\placefigure{20000_lc}

\begin{figure}\figurenum{12}\epsscale{0.9}
\plotfiddle{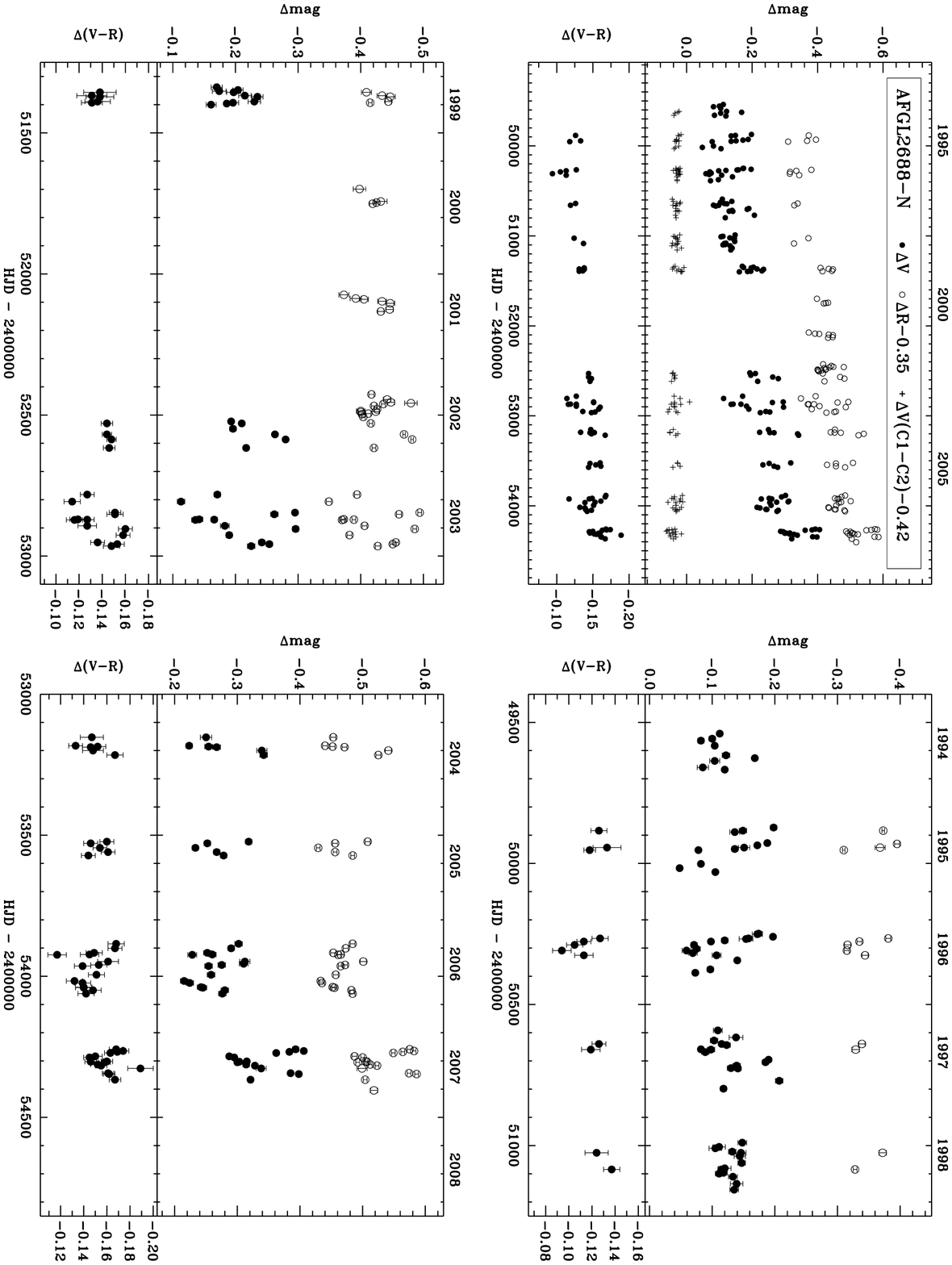}{0.0in}{+180}{400}{500}{0}{0}
\caption{Plot showing the differential light and color curves of AFGL 2688-N, plotted similar to Figure~\ref{22223_lc}.    \label{2688N_lc}}\epsscale{1.0}
\end{figure}

\begin{figure}\figurenum{13}\epsscale{0.9}
\plotfiddle{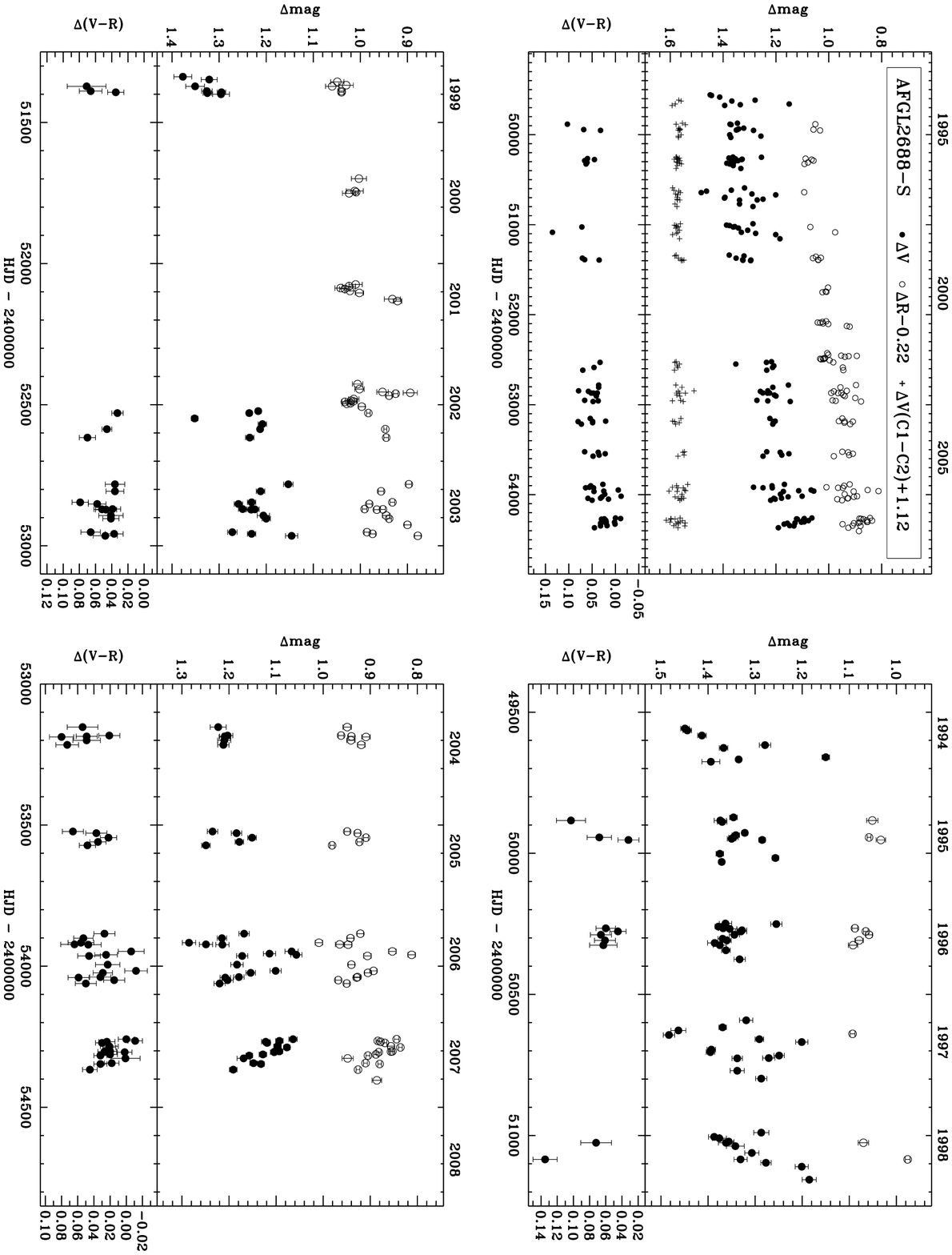}{0.0in}{+180}{400}{500}{0}{0}
\caption{Plot showing the differential light and color curves of AFGL 2688-S, plotted similar to Figure~\ref{22223_lc}.   \label{2688S_lc}}\epsscale{1.0}
\end{figure}

{\it AFGL 2688} (Figures~\ref{2688N_lc}, \ref{2688S_lc}).$-$  
This is the well-know bipolar PPN the Egg
Nebula and both lobes can be seen and measured separately.  
The southern (S) lobe is fainter by $\sim$1.5 mag.  
In carrying out aperture photometry, in order to avoid contamination by the northern (N)
lobe, a smaller aperture (7.2$\arcsec$ diameter) was used.  
Both lobes show a general trend of increasing brightness over the 14 seasons of
observation.  
Both lobes gradually increase in brightness by $-$0.22 mag (V) and get
bluer by $-$0.03 mag (V$-$R).
The N lobe also shows a cyclical variation within a season, with a 
peak-to-peak range varying between 0.04 and 0.18 mag (V) and averaging
$\sim$0.10 mag.  The variation in R is similar but slightly smaller.  
No obvious periodicity stands out from visual inspection
of the light curve.  The color is slightly redder when fainter within a season, 
reaching a maximum variation of 0.05.
The data from the fainter S lobe are less precise.  They also show a
cyclical variation that appears to be over a somewhat larger range 
of up to $\sim$0.30 mag (V) and $\sim$0.07 mag (V$-$R) in a season,
However, the variations in the two lobes do not appear to be correlated. 

\placefigure{2688N_lc}
\placefigure{2688S_lc}

\subsection{Trends in the Light Curves}
\label{lc_trends}

Several of the objects also show long-term trends in brightness over the observing interval.
IRAS 22223+4327 decreased in V brightness by $\sim$0.16 mag in 14 years.
IRAS 04296+3429 showed a monotonic increase in V brightness of 
$\sim$0.10 mag over 13 years. 
AFGL 2688, in both the N and S lobes, increased in V brightness by 
$\sim$0.22 mag and got bluer by $\sim$0.03 mag over 14 years.
This can be compared with an older, long-term study of the annual mean 
blue light curve of AFGL 2688 (N lobe) based on photographic plates.
This showed a general brightening from 1920 to 1958 of 
$\sim$2.0 mag followed by a more 
constant brightness through 1975 \citep{got76}.  Within a season a large spread 
in brightness was seen, indicating seasonal variation.  This rate of increase in 
brightness of the annual mean blue light curve of AFGL 2688, 
$\sim$0.05 mag yr$^{-1}$ (1920$-$1958)
or $\sim$0.04 mag yr$^{-1}$ (1920$-$1975), is three times as great as the rate that
we find of $\sim$0.015 mag yr$^{-1}$.
It is more similar to the smaller rate of increase seen in the photographic light curve
from 1968 to 1975.
IRAS 05341+0852 displayed a brightness increase beginning in 2005 that 
amounted to $\sim$0.09 mag (V) in four years.
IRAS 19500$-$1709 showed a sudden drop in V brightness of 0.12 mag in 1999
followed by a gradual increase over five years to a level $\sim$0.07 mag brighter than the 
initial value.   
It also became redder by 0.02 mag when it suddenly became fainter, 
but gradually became bluer by 0.04 mag as it subsequently brightened. 

\begin{figure}\figurenum{14}
\plotone{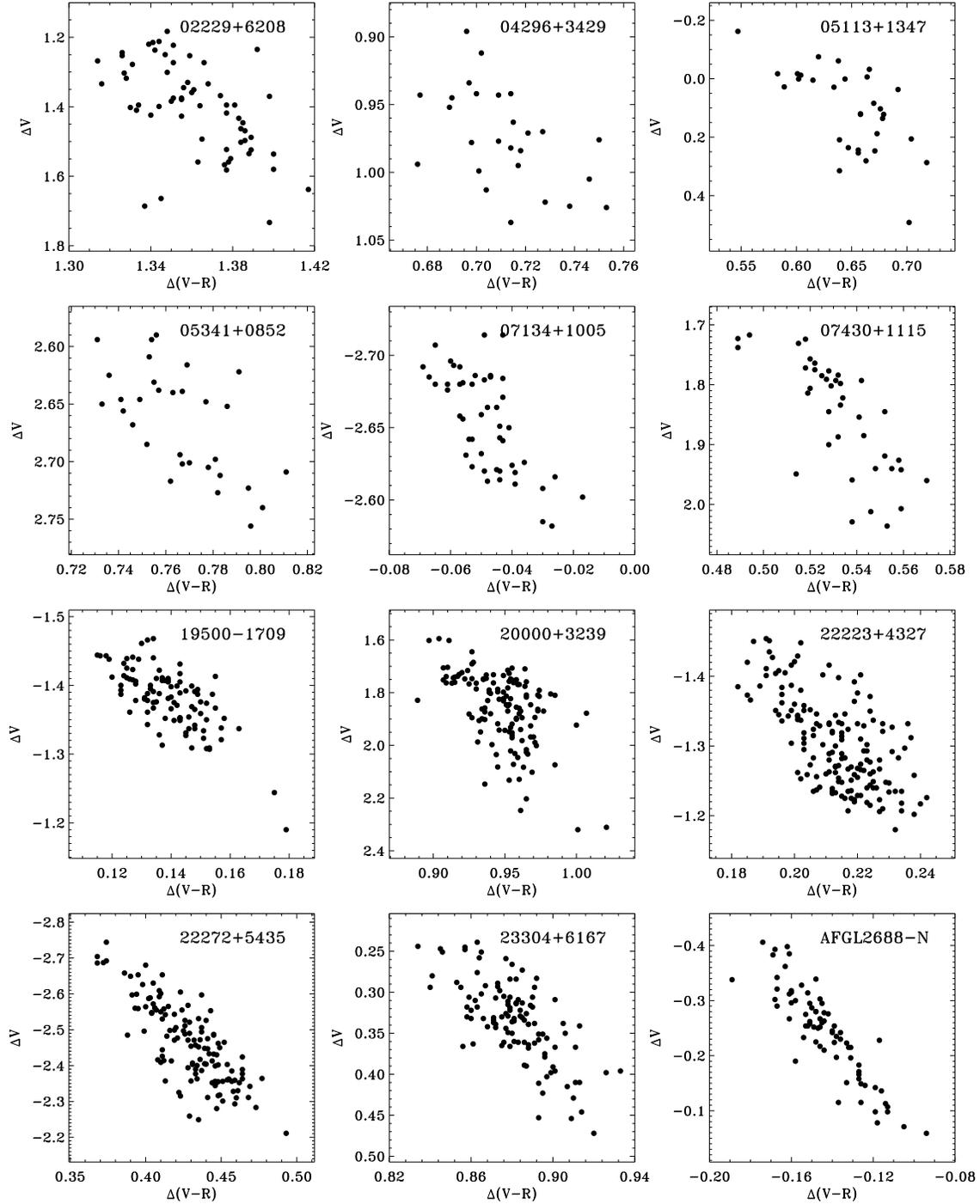}\caption{Plot showing the change in color with change in brightness for each of the targets.  A clear trend is seen for each of the PPNs, with the object being redder when fainter. \label{color-plots}}
\end{figure}

All of the objects show a general correlation between 
brightness and color, being redder when fainter. 
This can be seen quantitatively in a plot of differential color, 
$\Delta$(V$-$R) versus
differential magnitude, $\Delta$V.  These are shown in Figure~\ref{color-plots}, with a
panel for each object except AFGL 2688-S.  AFGL 2688-S shows a similar range in 
brightness and color as AFGL 2688-N, but with more scatter in the observations.  
However, the fainter S lobe is redder than the brighter N lobe by 
$\Delta$(V$-$R) $\approx$ 0.18; this is presumably
due to the greater extinction that causes the lobe to appear fainter.
While for almost all of these PPNs, this correlation between color and brightness
tracks the cyclical variations in the objects, this is not the case for
IRAS 19500$-$1709 and AFGL 2688-N and 2688-S;
in these objects most of the color change is associated with the long-term, seasonal light curve 
changes documented above and not the cyclical variations seen within a season.
It can be seen that the objects with the largest range in color, IRAS 05113+1347, 
20000+3239, 22272+5435, and 02229+6208, are the ones with the largest range in brightness.

\placefigure{color-plots}

\section{PERIOD ANALYSIS AND RESULTS}

\subsection{Period Search Methods}

The main goal of the program was to search for periodicity in the 
observed light curves.  This was investigated using several 
programs commonly used to search for periods in unequally spaced 
data.  We began with the CLEAN algorithm, which is based on a Fourier
transform \citep{rob87}.  This was used to investigate a dominant period 
in the data sets.
We further investigated the variability using the
period analysis program Period04 \citep{lenz05}.
This latter program also uses a Fourier analysis, fits the variations 
with sine curves of calculated amplitudes, and can easily be 
used to search subsets of the data and to investigate multiple periods.

For the PPN light curves that show a general trend in their seasonal 
light curves, such as a monotonic decrease (e.g., IRAS 22223) or
a sharp change (e.g., IRAS 19500), we removed the trend by normalizing
the seasonal light curves to the average value for that season. 
For others that appear to show changes in the mean light level from season to season, 
we also investigated the effect on the period search of normalizing the data to the 
mean level.

In the analysis of the data, we investigated the V, R, 
and (V$-$R) light curves separately.  
All of the objects show similar but not necessarily identical periods 
between the V and R light curves and with the 
(V$-$R) color curves where the data are of good precision.  
These differences likely arise from two related effects.
First, the data were not all taken in the same nights.  
The initial observations were only with the V filter and the use
of the R filter was introduced only gradually over the first several
years.  Then for approximately 2.5 years there is an absence of V data and
only R data are available due to a problem with the V filter.  Only since 2002 Aug 22 
($\sim$2002.6) have the V and R observations been consistently observed on the same
dates. The V data thus span a longer time span and they are more in number
but they are missing for about two and a half years.
Second, the period may be changing with time, and since the
V data are much more common in the early years, the resulting 
periods might be expected to differ.  
These differences are documented below for the individual sources. 
To investigate these differences further, for the objects with the most data, 
we also analyzed the periods in three subsets of the data: 
(a) the 1994$-$1999 V light curve, 
(b) the 2002.6$-$2008 V light curve, and
(c) the 2002.6$-$2008 R light curve.
These results suggest that, at least for a few objects, the period is changing slightly, by a few percent.

\begin{figure}\figurenum{15}\epsscale{0.8}
\plotone{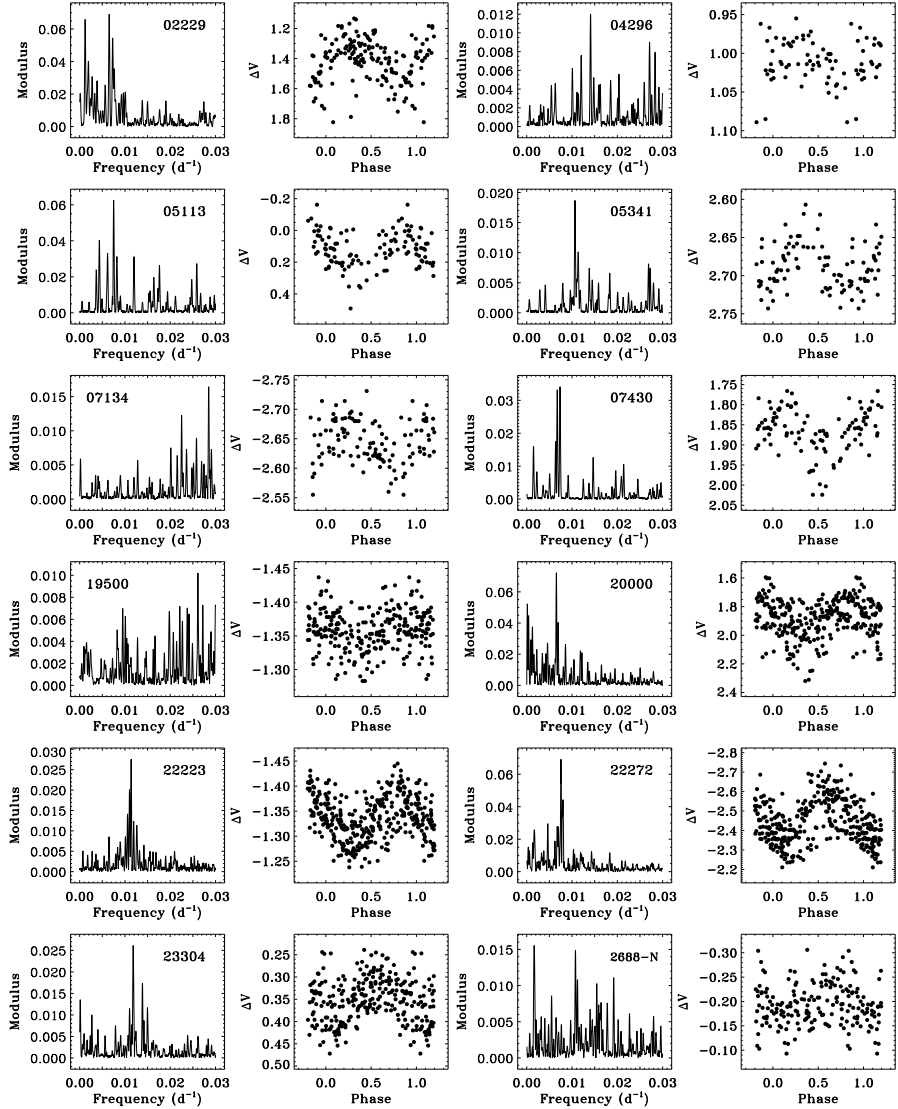}\caption{Plot showing the frequency spectrum and phased V light curve for each of the targets.  There is generally one strong peak in the frequency spectrum from which we derived the period.  The periods used are the first ones listed in Table~\ref{P_results} under the CLEAN analysis for this filter,
and the 0.00 phase is based on the time of the first data point.  The normalized light curves were used for the objects with large season-to-season trends: IRAS 04296+3429, 05341+0852, 07430+1115, 19500$-$1709, 22223+4327, and AFGL 2688-N lobe. \label{freqspec_V}}
\end{figure}

\begin{figure}\figurenum{16}\epsscale{0.8}
\plotone{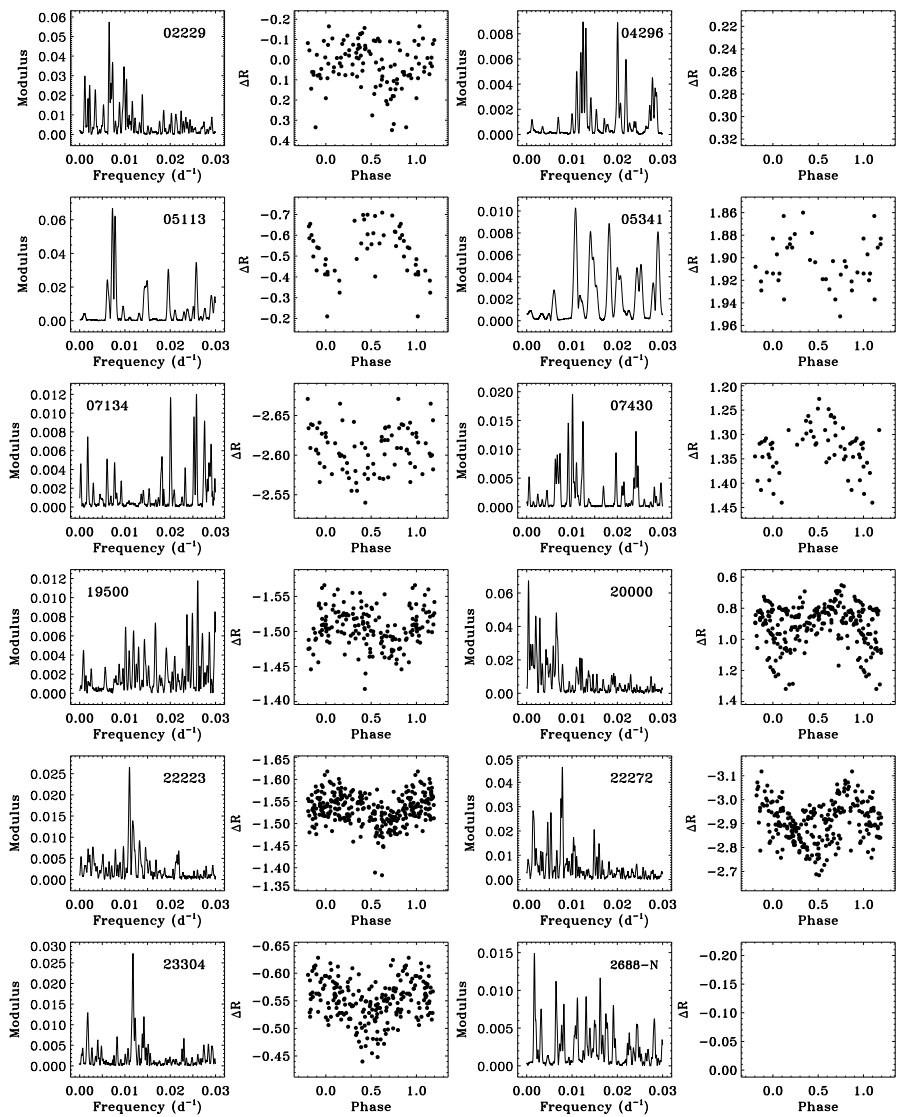}\caption{Plot showing the frequency spectrum and phased R light curve for each of the targets, similar to Fig.~\ref{freqspec_V}.  
IRAS 04296+3429 and AFGL 2688-N had no dominant peak, so no phased light curve was plotted. \label{freqspec_R}}
\end{figure}

\begin{figure}\figurenum{17}\epsscale{0.8}
\plotone{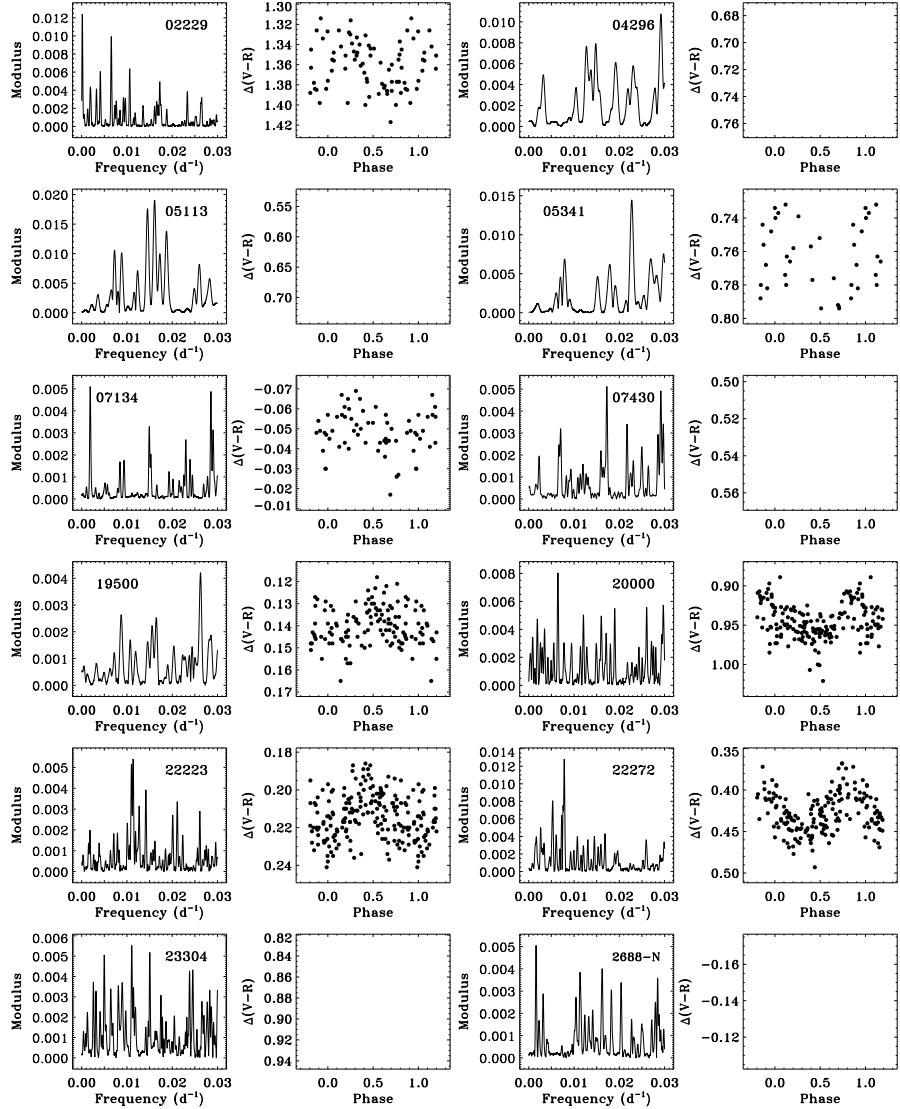}\caption{Plot showing the frequency spectrum and phased (V$-$R) light curve for each of the targets, similar to Fig.~\ref{freqspec_V}.
For the objects with no dominant peak, no phased light curve was plotted. \label{freqspec_VR}}
\end{figure}

The frequency spectrum from the CLEAN analysis and the phased 
light curve based on the dominant peak in the V, R, and (V$-$R) data (where applicable) are 
shown for each of the PPNs in Figures~\ref{freqspec_V},~\ref{freqspec_R}, and ~\ref{freqspec_VR}.
To determine the uncertainties ($\sigma$) in the calculated period values, we employed the 
least-squares fitting function in Period04.  
The individual analyses are discussed below.
In all cases, the results of the CLEAN and Period04 analyses agree
to within 2$\sigma$.  An overly conservative estimate of the uncertainty 
of the calculated period could be derived from the full-width, half-maximum of the period peak.  
This was examined in several cases and found to have a value of $\sim$10$-$20 times the 
least-squares uncertainty ($\sigma$).  In the following discussion, we have listed the period values to the nearest day.  The formal periods and uncertainties are listed in Table~\ref{P_results}.

\placefigure{freqspec_V}
\placefigure{freqspec_R}
\placefigure{freqspec_VR}

\placetable{P_results}

\subsection{Results for Individual Objects}

{\it IRAS 22223+4327}.$-$
The monotonically decreasing brightness trend was
first removed.  The period analyses revealed very clear, strong peaks
in the frequency. 
The V data show a strong peak at 
P = 88 d, with a secondary peak at 91 d; 
the R data show a strong peak at 91 d, and the (V$-$R) data 
show peaks at 88 d and 91 d.  
However, when we examine the data in subsets, 
for the 1994$-$1999 data, P(V) = 89 d and for the 2002$-$2008 data,
both the V and R data are similar with two equally strong frequency peaks at
P(V,R) = 91 and 86 d. 
Thus the data show the period to be in the range 86$-$91 d,
with evidence of multiple periods.
This encompasses the value of $\sim$90 days determined by \citet{ark03} 
based on UBV observations made over four seasons (1999$-$2002).

{\it IRAS 22272+5435}.$-$
The period analyses show a very strong peak in each light curve, 
with P(V) = 132 d, changing to P(V) = 131 d if we normalize the light curve 
for the slightly different seasonal means, and a second double peak at 124.9/127.7 d.  
For the R data, P(R) = 127 d with a weaker peak at 132 d; 
the (V$-$R) data show a dominant period of 127 d.  
If we investigate this in subsets, 
for the 1994$-$1999 data, P(V) = 133 d, and for the 2002$-$2008 data,
both the V and R data yield the same period P(V,R) = 128 d. 
Thus for the object, the
data show the period to be in the range 127$-$133 d, 
with evidence for multiple or changing periods.
These values differ significantly from the
periods of 207 days and 145 days (secondary period) determined by 
\citet{ark00} based on 72 observations in each of the UBV filters 
made over nine seasons (1991$-$1999).  
This demonstrates the intrinsic variability of the light curves of 
these objects and the need for long-term studies to accurately determine
their periods.
Recently \citet{zacs09} determined a period of 131 d based on radial velocity
observations, a value in good agreement with ours.
These additional independent data sets will be combined in our follow-up 
study of this object (Paper II).

{\it IRAS 23304+6147}.$-$
The period analyses result in P(V) = 85 d and P(R) = 85 d,
with a secondary period of 72 d (V) and 70 d (R). 
The (V$-$R) data set shows no dominant period.
Examining the data in subsets yields generally similar results: 
P(V) = 84 and 72 d for the 1994$-$1999 V light curve, 
P(R) = 81 d for the 2002$-$2008 R light curve, but
P(V) = 67 d with low amplitude for the 2002$-$2008 V light curve.
The reason for this deviant period value for the last data set is not clear
(it does show a secondary period of 91 d), 
but the rest of the data give a consistent value for the dominant
period $\sim$85 d for this object.

{\it IRAS Z02229+6208}.$-$
The period analyses result in similar periods in the two colors
of P(V, R) = 153 d, with the same period when normalized.
A longer period of $\sim$820 d appears in the V data, 
although it is much less significant when the data are normalized.
A weaker peak of 137 d also appears in the data.
The (V$-$R) color curve has a consistent peak at 153 d;
the long period peak in the frequency spectrum is due to the trend in the
color data and goes away when the color curve is normalized .
An examination of the data in subsets shows periods of P(V) = 152 and 820 d
for the 1994$-$1999 V light curve and P(V, R) = 136 d for the 2002.6$-$2008
light curves, but these are less certain due to the small number of data points
($<$40) in each subset.
Thus the dominant period for the system appears to be P = 153 d.
There is no evidence for a period of twice this value, as might be the case for
an RV Tau star with alternate deep and shallow minima.

{\it IRAS 04296+3429}.$-$
The period analyses of the normalized data show 
no dominant period, but the strongest peak is 
at P(V) = 71 d with no strong peak in the smaller R and (V$-$R) data sets.
Thus P = 71 d should be regarded as a tentative value.

{\it IRAS 05113+1347}.$-$
The period analyses result in a period of 133 d in V and 
138 and 128d in R.  There is no dominant period in the (V$-$R) data set.
The data from 2002.6$-$2008 alone yield P = 136 d in both V and R, 
with a larger amplitude than when the entire data sets are analyzed,
but the sample size is small ($<$30).
Thus the period appears to be in the range $\sim$133 d.

{\it IRAS 05341+0852}.$-$ 
The period analyses were carried out on the normalized data.  
The V data gives a clear period at 94 d.  The R data are very few in number most years, 
which makes normalization uncertain,
except for the four-year span from 2003$-$2006.
An examination of the R data from 2003$-$2006 alone indicates P = 93 d, 
based on 26 data points.  The small (V$-$R) data set possesses no dominant 
peak in the frequency spectrum.

{\it IRAS 07134+1005}.$-$
The period analyses show peaks at P(V)  = 35 d and 44 d,
P(R) = 39 and 50 d, and P(V$-$R) = 35 d (along with a longer period of 
528 d), but none of them is very strong.
If we normalize the data to the mean light level, then P(V) = 39 d
and P(R) = 39 d.
Analyzing the data in subsets results in 
P(V) = 44 and 39 d for 1995$-$1999, 
P(V) = 35 d for 2002$-$2008, and
P(R) = 39 and 35 d for 2002$-$2008, 
but the number of data points is small for each, 49, 40, and 39,
respectively.
It seems likely for this object that there are 
either several relatively closely spaced periods or that the period is
switching from one value to another.  The period value is found to be
in the range 35$-$44 d, with a most probable value of P $\approx$ 39 d.

This bright PPN has previously been studied for periodic variability.
\citet{bar00} found a main period of 36.7 d based on 89 radial velocities 
obtained over eight years and 36.9 d based on 87 photometric measurements 
made over seven seasons (1988 to 1996, with most in the first four seasons).
These values are consistent with our results, and these data and ours will 
be combined in our subsequent detailed study of this object (Paper III).

{\it IRAS 07430+1115}.$-$
The period analyses were carried out on the normalized data.
These resulted in two nearly equally probable periods in V, 136 d and 148 d, 
with the former slightly preferred based on the appearance of the resulting 
phased light curve.
The R data has its strongest peak at 99 d, with the second strongest at 136 d;
the latter one agrees with the V peak and gives a good phased light curve.  
There is no strong peak in the (V$-$R) analysis.
Thus the best fit to the data, based primarily on the more numerous V data, 
indicates that P = 136 d, with a secondary peak at 148 d.
We are continuing observations of this target to better determine the period.

{\it IRAS 19500$-$1709}.$-$
The period analyses was carried out on the normalized data 
sets and indicated a period of 38 d in V, R, and (V$-$R). 
Other, weaker periods appear to also be present in the data.
When the data were examined in subsets, we found the following:
P(V) = 51 and 37 d for the interval 1994$-$1999;
P(V) = 42 and 38 d for the interval 2002$-$2007; and
P(R) = 38 d for the interval 2002$-$2007.
The period of variation for this object appears to be less stable than 
for the longer-period objects in this study.  
We conclude that the most dominant value of the period during our
observing interval is 38 d, but there appear to be other periodicities in the data.

\citet{ark00}, based on their 1993$-$1999 UBV light curves study,
found periods of 97.4, 53.1, and 34.9 d, none of which were very strong. 
In our detailed study of the light and velocity of this object (Paper III), we will  
combine our data with that of \citet{ark00} and with some additional unpublished
photometric data in an effort to better understand the pulsational properties of
this object.

{\it IRAS 20000+3239}.$-$
The light curve shows distinct variations in seasonal brightness.
We began by analyzing the observed, un-normalized data.
The period analyses indicate a well-determined period 
in V at 153 d.
The R data also showed a period at 153 d; in addition, the frequency spectrum also
shows a period of about 7 years from an attempt to fit the seasonal brightness variations.
The (V$-$R) color curve indicating P = 155 d.  
The V, R, and (V$-$R) light curve periods are similar when normalized, 153, 
152, and 156 d, respectively.
Examining the 2002$-$2008 light curves separately yields the same period
values, but with larger amplitudes than found for the entire data sets.
There is no evidence for a period twice as long.
Thus the light curves indicate that P $\simeq$ 153 d.

{\it AFGL 2688}.$-$
We investigated the presence of a period in the brighter, North lobe of this
extended bipolar object.  The light from the central star is complete obscured.
The period analyses suggest a long period of about 600 d
and slightly weaker periods of 93 and 87 d in V and 62 and 89 d in R, following the removal
of the long-term brightening trend in the data.
The $\sim$600 d periodicity shows up on the (V$-$R) frequency spectrum, along with 
several other weak peaks.
We take the best estimate of the pulsation period of the star to be $\sim$90 d.
This object bears continued monitoring to verify the reality of this and the longer period.

Note that the visible light that we are receiving from this object is light
scattered toward us from the lobes.
This makes the detection of a periodicity in the light of the central 
star more difficult since we are measuring the light of the object integrated
over an area of the lobe.  The {\it HST} image of the object \citep{sah98} 
shows the brighter portion of the North lobe to have a radial size of about 5$\arcsec$, 
and a measurement of the size of the lobe in our images shows the FWHM to be 
about 5$\arcsec$.
Typical distance estimates range from  400 to 1500 pc.
At an estimated distance of 1 kpc,  5$\arcsec$ subtends an size of about 30 light days.
If we adopt the recent value of 420 pc derived from differential proper motion based 
on {\it HST} images \citep{ueta06}, then the bright part of the lobe subtends a size
of $\sim$12 light days.
In practice then, this light spread is convolved with the periodic variation from
the star and thus diminishes the emitted profile of the light variation,
making a periodic light variation less than $\sim$100 d difficult to measure.

\subsection{Evidence for Period Changes}
\label{p_evln}

Evidence for period changes was investigated in three of the targets possessing the 
most data and with relatively well-determined periods.
We compared the period determined for the 1994$-$1999 interval with that determined 
for the 2002$-$2008 interval.
For IRAS 22223+4327, the period changed from 89 d to two equally strong
periods of 86.5 and 91.5 d, which have an average value of about 89 d.
For IRAS 22272+5435, the period appeared to decrease from 133 to 128 d and 
for IRAS 23304+6147, the period appeared to decrease from 84 to perhaps 81 d.
These latter two suggest perhaps a decrease of $\sim$4$\%$ between these two time intervals.
However, these results are only tentative since the time interval is short, on the order of
a decade, the sample is small, and these objects show evidence of multiple 
periods.

We will investigate this question of period change further for IRAS 22223+4327 and
22272+5435 in Paper II, where we include the results of other published photometric 
data for these two sources.  We will also there investigate the evidence for multiple 
periods in these two systems that can lead to the appearance of a beat period.
In addition, we are continuing the observations of all three of these  
PPNs to extend the time baseline.

\section{DISCUSSION}

The targets in this study are all post-AGB objects in transition between the 
AGB and PN phases in the evolution of intermediate-mass stars.
It is known that AGB stars vary in brightness with large amplitudes due to 
pulsations as they evolve from Miras to long-period variables to OH/IR stars, 
losing mass at an increasingly high rate \citep{olof04, hri05}. 
Periods for these range from 100 to 2000 d.
Once in the post-AGB phase, such stars evolve at constant luminosity
toward higher temperatures.
This is the state of our targets, which are evolving to higher temperatures 
and shorter periods, presently with spectral types ranging from $\sim$G8 to F3
and periods ranging from 153 to $\sim$38 days.
Most of these PPNs have model atmosphere analyses based on high-resolution spectroscopy, 
from which effective temperatures ({\it T}$_{\rm eff}$) have been determined. 
These results are all summarized in Table~\ref{results}. 
There exists a clear trend of period with spectral type; 
the coolest PPNs have a spectral type of $\sim$G8, an
effective temperature of 5250$-$5500 K, and a pulsation period of $\sim$150 d.
With a {\it T}$_{\rm eff}$ range of 5250$-$8000 K, these objects span the temperature range of the instability strip for luminous variable stars and perhaps extend even a little hotter.
This suggests that these objects also vary due to radial pulsation driven by the
{\it $\kappa$}-mechanism \citep{ost96} operating in the helium partial ionization zone.

\placetable{results}

In respect to spectral type and period, the PPNs in this study bear some resemblance to the RV Tau stars, which are also variable stars thought to be post-AGB stars of F$-$K spectral types and luminosity
class of Ia-II.  
The RV Tau stars show alternately deep and shallow minima, 
with periods of approximately 15 to 75 d (or twice that between deep minima),  
and {\it V} amplitudes generally between 1 and 2 mag, 
although some have even larger amplitudes \citep{wahl93, per07}.
A few RV Tau stars with {\it V} amplitudes as low as 0.1 mag 
have been observed \citep{pol96}.
It has been found that RV Tau stars share a period$-$luminosity
relationship with Type II Cepheids, as seen from studies in the LMC \citep{alc98}.
In the LMC study, it is also found that the longer period RV Tau
stars have smaller amplitudes, the opposite of what we find in our PPN sample
(see below) and that there is no relationship between period and color,
while in our PPNs there exists a clear relationship between period and spectral 
type (see Table~\ref{results}).
The lack of distances and thus luminosities for these PPNs prevents
us from investigating directly a period$-$luminosity relationship;
however, the evidence in hand points more strongly to a period$-$temperature
relationship based on the evolution of these objects
across the HR diagram.
While the RV Tau 
stars also possess infrared excesses, 
they are generally much less than that seen in these PPNs, 
for which about half of the flux is detected in the mid-infrared.  
This difference in circumstellar material points to a difference 
in the initial masses, with these PPNs evolving from stars of higher 
initial mass than the RV Tau variables, which are thought to evolve from
low mass stars.
Thus these PPNs, in general, tend to have longer periods and smaller
amplitudes than the RV Tau stars, without (in most cases) 
alternating differences in the depth of minima.
Recent radial velocity studies indicate that a significant fraction of RV Tau 
stars (and other dusty post-AGB stars) may be binaries in which the circumstellar 
dust is maintained by the gravitational interaction of the companion 
\citep{vanwin03,vanwin07}.
None of the PPNs have thus far been found to be radial velocity binaries \citep{hri09e}.
Future studies of an enlarged sample of PPNs can investigate further the
similarities and differences with the pulsational light curves of RV Tau stars 
and other dusty post-AGB stars.

This observed upper limit to the PPN periods of $\sim$150 d bears upon 
models of the post-AGB evolution of intermediate-mass stars.
Post-AGB stars evolve toward higher temperatures and the PN regime at
rates that depend upon their envelope mass relative to their core mass.
Evolutionary calculations show that the evolution from the tip of the AGB
to the beginning of the PPN phase,
from T$_{\rm eff}$ of $\sim$3000 to $\sim$5000 K, 
will take much too long ($\sim$10$^5$ years depending upon the core mass)
unless there is significant post-AGB mass loss until 
T$_{\rm eff}$ $\approx$ 5000 K \citep[eg.,][]{sch83}.
Various prescriptions for such ongoing mass loss have been applied \citep{vas94,blo95}.
\citet{blo95}, in his models of post-AGB evolution, made the assumption that the
extensive mass loss rate does not simply terminate abruptly at the end of the AGB 
phase, but rather that it decreases gradually in proportion to the decreasing radial 
pulsation period of the star.
(This dependence of mass loss on the pulsation period is quite reasonable since 
pulsation seems to be the mechanism driving mass loss through dust formation
on the AGB \citep{bow88}.)
In most of his models, Bl\"{o}cker assumed that the extensive mass loss rate 
began decreasing when the star had a pulsation period of 100 d
and decreased down to a low Reimers mass-loss rate as the period lowered to 50 d.  
He found from these models that the time scale for this evolution
was relatively rapid, on the order of a several hundred years or less.  
This initial assumption was changed for one of his published models, 
in which he started the decrease in 
mass loss rate with the longer pulsation period of 150 d and decreased the 
rate as the period lowered to 100 d.
In this later case, Bl\"{o}cker found that the stellar evolution during this transition
in mass loss rate happened much more slowly, 
$\sim$2800 yr when beginning with P = 150 d as compared to $\sim$800 yr 
when beginning with P = 100 d (with M$_{\rm ZAMS}$ = 3.0 M$_\sun$ and 
M$_{\rm core}$ = 0.625 M$_\sun$).\footnote{The \cite{blo95} models 
with M$_{\rm core}$ = 0.604, 0.696, 0.836, and 0.940 
M$_{\sun}$ all had even shorter time scales of $\le$ 160 years for the decrease in 
mass loss rate, based on the mass loss decrease occurring between pulsation periods
of 100 d and 50 d.}
Thus this change to a longer initial period for the beginning of the decrease in 
extensive mass loss rate has 
the consequence of significantly slowing down the evolution of the star from the
tip of the AGB to the start of the PPN phase.

The pulsation periods found in this present study can help inform these 
stellar evolution models.  Clearly the peak of the extensive mass loss has ended 
by the time the pulsation period is 150 d and the effective temperature is 5250 K.  
The SEDs of these PPNs attest to this, for, with the exception of AFGL 2688, 
the central stars are not highly obscured and the double-peaked 
SEDs indicate that the circumstellar shell is detached.
Dust temperatures for the shells are T$_{\rm d}$ $\approx$ 200 K, well below the
dust condensation temperature for amorphous carbon of $\ge$1000 K 
\citep[with a recent value as high as $\ge$1600 K;][]{speck09}.
About half of the PPNs in this sample have been imaged in the mid-infrared and
their detached dust shells resolved, including IRAS 22272+5435 \citep{ueta01}
with a the G5 spectral type.
Based on Bl\"{o}cker's model, this longer initial pulsation period would imply that the 
transitional time from the end of the extensive AGB mass loss to the beginning of the 
ongoing low mass loss rate is a few thousand years rather than a few hundred.
The models of the SEDs of these PPNs support this, as dynamical
times scales of 1000$-$2000 years have been found in most cases for the ages of the
densest region of the dust shells \citep[for L=8000 L$_\sun$]{hri09,hri00}.  
These observed SEDs agree with the model SEDs determined from hydrodynamic,
radiative transfer models of the evolution of the central star and the dust shell by
\citet{stef98} and \citet{sch07}, although the time scales of their models are somewhat
shorter, 500$-$700 years.
While some of these observed PPNs do show a suggestion of some warm dust in their SEDs,
which could be evidence for more recent, lower rate mass loss \citep{hri09},
one of the cases, IRAS 05113+1347, has P = 133 d, again supporting the
fact that the decline from the intensive mass loss begins at P $>$ 100 d. 
An additional support for the longer initial evolution time scales is the fact that 
most of the 12 objects in this study have T$_{\rm eff}$ $<$
7000 K (9/12) rather than T$_{\rm eff}$ $\ge$ 7000 K (3/12),
suggesting a slow initial evolution, since there should not be an observational bias  
against detecting the hotter C-rich PPNs.
There are four known additional hotter C-rich PPNs, IRAS 01005$-$7910 
(B0e)\footnote{We have observed IRAS 01005$-$7910 and found 
it to vary on a short time scale of less than a few days.
None of the others have published light curve studies.},
IRAS 16594$-$4656 (B7e), and two recently discovered by
 \citet{rey07} with spectral types of A8 and A9 and T$_{\rm eff}$ of 7750 and 8000 K.  
Even including these does not change the result that a majority have cool
central stars.  
This result suggests that the actual time scales to evolve to the central stars of PNs is 
significantly longer than determined in the Bl\"{o}cker models in which 
evolution with extensive mass loss continues until the pulsation period P = 100 d and then decreases.
However, the \citet{blo95} model in which mass loss begins to decrease at P = 150 d 
(rather than 100 d) has a evolutionary time scale of $\sim$7,000 yr 
for the PPN phase, from the end of the decreasing AGB mass loss 
to the beginning of the PN phase (in addition to the 2,800 yr from the peak to 
the end of the decreasing AGB mass loss rate).
This is too long a value for the PPN phase and suggests that the post$-$AGB 
mass loss rate is higher than the Reimers rate assumed in the Bl\"{o}cker study.

\begin{figure}\figurenum{18}\label{period-temp}
\plotfiddle{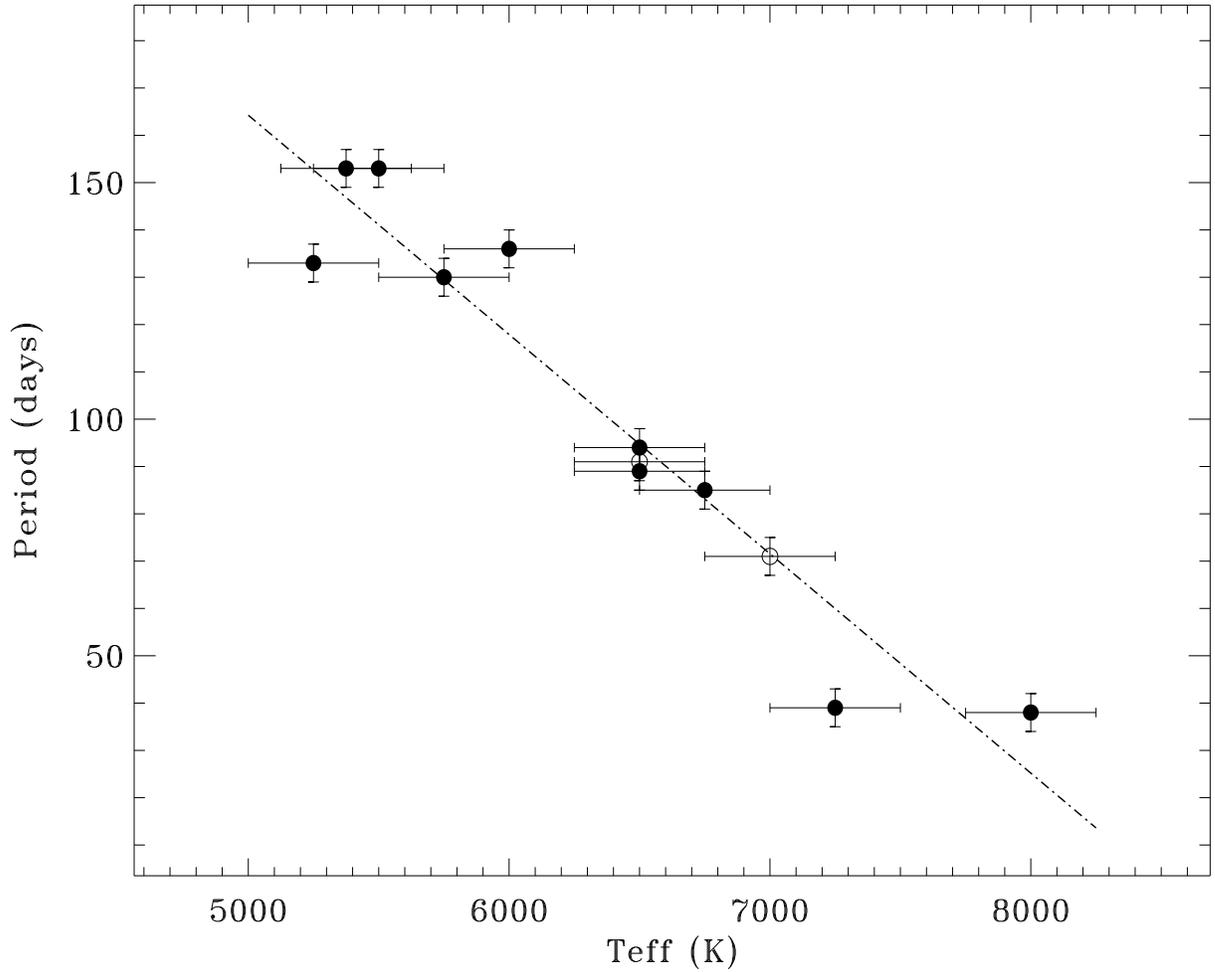}{0.0in}{+90}{400}{500}{0}{0}
\caption{Plot showing the pulsational period of each target compared to its temperature. A clear trend is seen for each of the PPNs, with the period being shorter when the object is hotter.  The uncertain period values for IRAS 04296+3429 and AFGL 2688 are shown as unfilled circles.  (We have assumed reasonable error bars of $\pm$250 K and $\pm$3 d.) }
\end{figure}

One would like to be able to empirically determine the evolution of {\it P} (and 
$T$$_{\rm eff}$) with time for individual PPNs, 
as has been attempted for some RV Tauri variables \citep{per91}. 
This would allow us to better constrain the models and the post-AGB 
mass loss rates.
From the present study, one can use the observed relationship between
period and temperature to show the average evolution of these two quantities.  
This is shown graphically in Figure~\ref{period-temp}.
An approximately linear relationship can be seen over the limited temperature 
range of 5250 to 8000 K covered in this study.
From the slope of this relationship, we can determine a rate of change in the 
pulsation period $\it P$ with $\it T$$_{\rm eff}$ of 
$\Delta$$\it P$/$\Delta$$\it T$$_{\rm eff}$ = $-$0.047 d/K.
While the actual evolution in time of an individual PPN will depend upon
M$_{\rm core}$  \citep[and M$_{\rm ZAMS}$; see][]{blo95} and the level 
of ongoing post-AGB mass loss, this gives a general result.
If we assume, for example, that these stars have completed the transition in
mass loss from the high AGB rates and are now losing mass at the much lower
Reimers rate, then we can couple this with one of the post-AGB evolution 
models to calculate the expected evolution in pulsation period.
We will use the results of \citet{stef98} following the end of the decreasing
mass loss from the AGB to the Reimers rate.  Although we showed above that 
time scale values from this model must be adjusted for the fact that the AGB rate 
appears to end earlier than assumed in the model, the model evolutionary rate 
should still be appropriate in the Reimers regime.
As stated by \citet{sch83}, ``the evolution depends only on the {\it current} 
values of $\dot M$ and is independent of earlier mass-loss phases, provided 
thermal equilibrium has not been destroyed.''
\citet{stef98} find, for a carbon-rich star with M$_{core}$ = 0.605 M$_{\sun}$, 
an evolution from $\it T$$_{\rm eff}$ of 6050 K to 8500 K occurs in 675 years, or a rate
of 3.6 K/yr.  Combining this with our empirically-determined average rate of change of
pulsation period with $\it T$$_{\rm eff}$ results in a rate of change in period of
$\Delta$$\it P$/$\Delta$$\it t$ = $-$0.17 d/yr or a change of 
$\Delta$$\it P$ = $-$2.4 d over the 14 year observing interval.
Such a change should be measurable with sufficient observations over a long enough
baseline.  
We found a tentative value for the rate of period change of this order for two of
the best observed cases (Section~\ref{p_evln}).
However, we caution that this is based on a relatively small temporal baseline and will need to be 
monitored for an additional decade or so before one can begin to feel confident
in the results.  Also, the results would be strengthened by the addition of more PPNs 
to the determination of the P$-$T$_{\rm eff}$ relationship.  
We are in the process of carrying out a similar light curve study of oxygen-rich PPNs.
Note that \citet{sas93} nicely outlined some of the general evolutionary aspects of 
pulsating post$-$AGB stars that are in good quantitative agreement with 
the results found in this study.
When we examine the P$-$T$_{\rm eff}$ relationship on a logarithmic scale, 
we find a slope of $-$3.5$\pm$0.4, within 
the range of $-$3.0 to $-$3.7 listed by Sasselov and close to the theoretical value of 
$-$3.72 based on the fundamental-mode pulsation in Miras \citep{ost86}.
 
\placefigure{period-temp}

\begin{figure} \figurenum{19}\epsscale{1.1} \label{delv-period-temp}
\rotatebox{90}{\plottwo{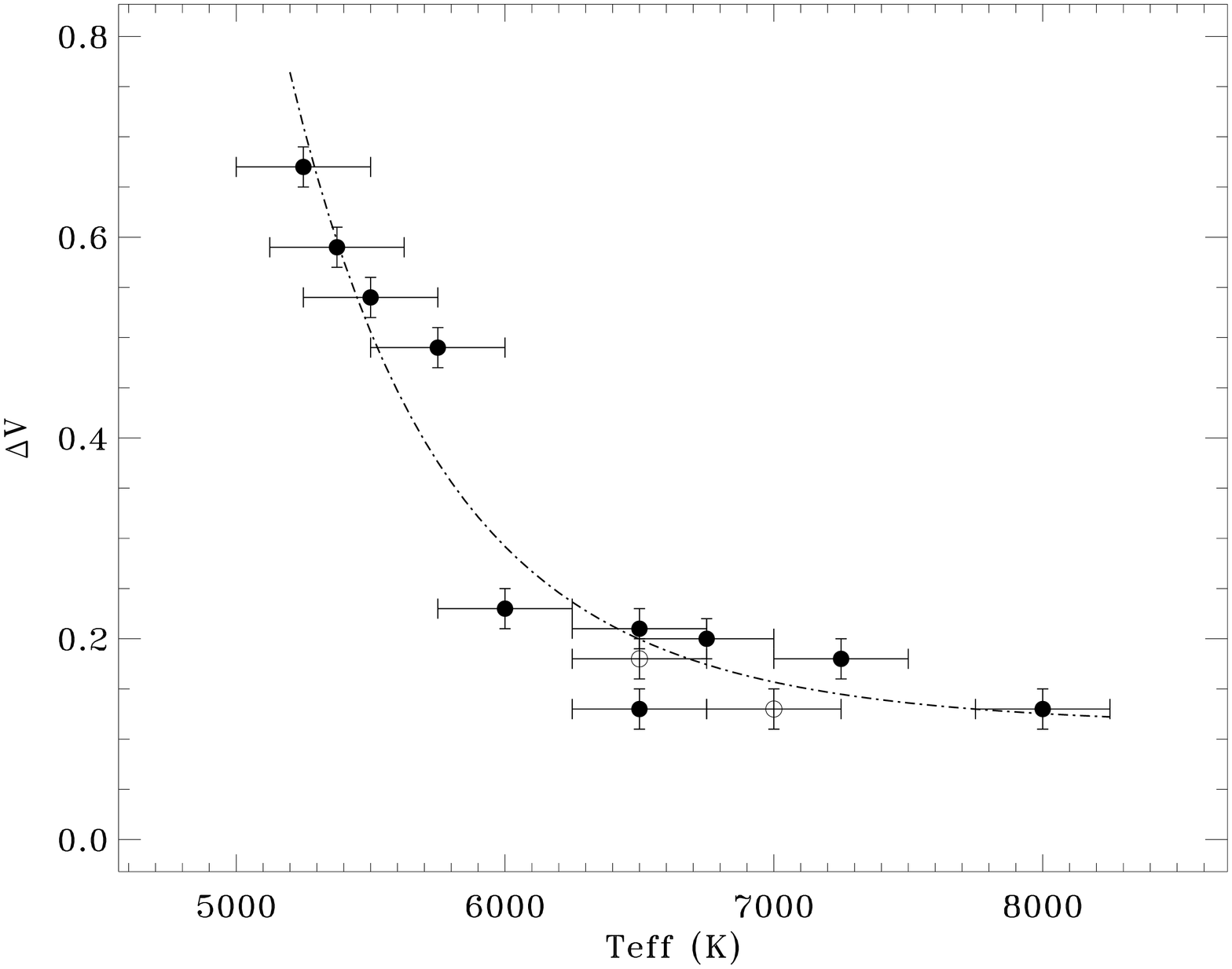}{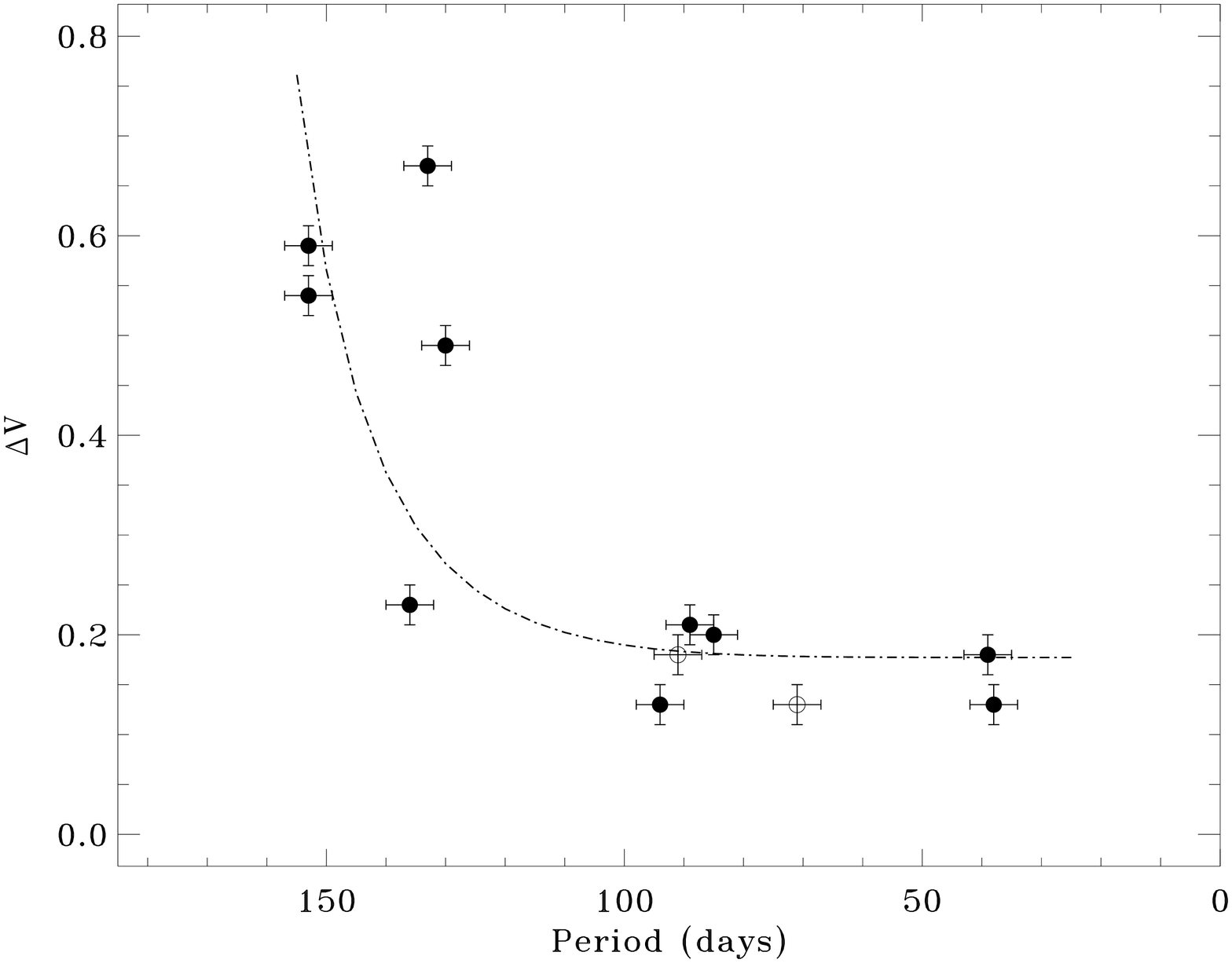}}
\caption{Plots showing the decrease in the amplitude of the variation ($\Delta$V) of each PPN as a function of its period and effective temperature.  The uncertain values for IRAS 04296+3429 and AFGL 2688 are shown as unfilled circles.  Dashed lines are added to show the trends.}
\end{figure}

The peak-to-peak range of the light variation also shows a trend with period.  
The PPNs with the longest periods also have the largest light curve variations.
Listed in Table~\ref{results} is the maximum seasonal variation in the
V band of each PPN.  
In Figure~\ref{delv-period-temp}, these maximum variations are seen to decrease 
approximately monotonically with decreasing period and with increasing effective temperature.  
This decrease in light curve variation with P and with {\it T}$_{\rm eff}$ is what 
one would expect to see as the atmosphere decreases in size while the star evolves 
at constant luminosity toward higher temperature.
Note that these maximum seasonal variations are much larger than the amplitudes determined
from the sine curve fitting to the long-term light curves; this is at least partly due 
to the multiple periods or varying amplitudes inherent in these pulsating stars.
The comparison of the light variations with pulsation periods shows a marked 
transition at period values of $\sim$120-130 d, below which the amplitudes are
low and in the range 0.13$-$0.23 mag, while at longer periods they are larger,
in the range of 0.5$-$0.7 mag.  
(IRAS 07430+1115 is the lone exception; it is cool (6000 K) with a long
period (136 d) but a small variation (0.23 mag).)
This may suggest that at periods of 120$-$130 d is when the mass loss 
rate has ended the transition from a non-abrupt, decreasing AGB rate to a 
lower post-AGB rate.  This appear to correspond to a temperature of $\sim$6000 K.
The larger light variations at the longer periods supports the idea that the 
pulsation is stronger and better able to drive mass loss at these periods.
Thus it may be that P $\approx$ 125 d and T$_{\rm eff}$ $\approx$ 6000 K
are the parameters where the decreasing AGB mass loss ends and a lower post-AGB
mass loss begins.

The observed values of pulsation period and brightness range can, in principle, be compared 
to pulsation models to both constrain the model parameters ({\it M, L, T$_{\rm eff}$}, 
metallicity)
and to help interpret the observed light (and velocity) curves.
\citet{fok94} has published a series of nonlinear pulsation models for cool post-AGB stars with
{\it M} = 0.6 M$_{\sun}$, {\it T$_{\rm eff}$} = 4600$-$5600 K, and {\it L} = 2000$-$8000 L$_{\sun}$.  His goal was to investigate the light curves of RV Tau variables.
These produced model light curves with shorter pulsation periods (by about a factor of 
one-half) than those observed for the 
cool (5250$-$5500 K) PPNs in this study and with amplitudes much larger ($\ge$ 1 mag)
than found in the PPNs.
In a later study, \citet{fok01} have presented a series of radiative, nonlinear pulsation models
of post-AGB stars with parameters in the following ranges:
{\it M} = 0.6$-$0.8 M$_{\sun}$, {\it T$_{\rm eff}$} = 5600$-$6300 K, and 
{\it L} = 4500$-$8000 L$_{\sun}$.
These model parameters were chosen in part to fit the observed light and velocity curves
of the PPN IRAS 07134+1005.  
The results of these studies are that the light curve periods and amplitudes decreased with increasing
temperature and also decreased with increasing mass and decreasing metallicity.
More model calculations, with the explicit inclusion of convection and producing larger periods, are needed to help interpret these new PPN light curves.

\placefigure{delv-period-temp}

The observed maximum changes in $\Delta$(V$-$R) attributable to temperature changes, 
excluding the changes associated with trends in the brightness, are listed in Table~\ref{results}.
These were transformed to changes in temperature based on the published temperatures and 
a color-temperature table of \citet[Table 15.7]{cox00}.  These $\Delta${\it T$_{\rm eff}$} values range from $\sim$300$-$750 K, with no correlation with {\it T$_{\rm eff}$}.  In some cases, such as IRAS Z02229+6208 and 07430+1115, these are likely lower limits since the R observations were few in the early years.
As noted earlier, the observed (V$-$R) colors are reddened by the interstellar and particularly circumstellar dust, and thus the
temperatures are derived from the spectroscopic analyses.

The observed long-term trends in the brightness of a few of the PPNs can most likely 
be ascribed to the effects of changes in the line-of-site circumstellar dust.  
IRAS 22223+4327 appears to get gradually fainter and IRAS 04296+3429 and AFGL 2688 
appear to get gradually brighter.  
AFGL 2688 shows the largest change in brightness and gets measurably bluer as it gets
brighter.  
IRAS 19500$-$1709 shows a rapid drop in brightness, during which time it gets
redder, and then a gradual rise in brightness during which it gets bluer.  This can be
explained by the sudden formation and more gradual dissipation of a dust cloud or
the passage of a circumstellar dust cloud with a more opaque leading edge.

\section{SUMMARY AND CONCLUSIONS}

In this study, we carried out a photometric V and R monitoring survey of 
12 carbon-rich PPNs over an interval of 14 years.  
This is the first systematic, long-term ($>$10 yr)  light curve study of a 
homogeneous group of PPNs.
The targets were all of spectral type F and G.
Below are the primary results of this study.

1. All of these PPNs varied in brightness and color, with the general
result that they are redder when fainter.

2. Periods of the variations were found for all of the targets (except perhaps AFGL 2688).
These range from $\sim$38 to 153 days, with a maximum V variation 
(peak-to-peak within a season) ranging from 0.13 to 0.67 mag.  
The objects with the later spectral types, mid- to late-G,
have the longer periods and the larger variations.

3. The objects show evidence for multiple periods and/or changing periods and/or 
changing amplitudes.  
Two of them have light curves that suggest a longer beat period (IRAS 22272+5435 
and 22223+4327) and two of them have light curves with a suggestion of intervals of 
alternating deeper and shallower minimum (IRAS 20000+3239 and Z02229+6208).
Nevertheless, all except the two with the shortest periods 
(IRAS 07134+1005 and 19500$-$1709) show a clearly dominant period in the
data set.  These results are consistent with the variation being due to pulsation and not 
binary interactions.

4. An approximately linear correlation was found between the period of variation and the
effective temperature over the temperature range observed in this study (5250$-$8000 K), 
with the shorter periods associated with higher temperatures.
The range of light variation is also seen to decrease approximately monotonically with temperature.

5. The presence of pulsation periods as long as 150 d in PPNs with detached shells implies that the 
intensive AGB mass loss had ended before the period had decreased to 150 d.
This has significant implications for modeling post-AGB evolution rates, since the rates depend upon the  level of post-AGB mass loss.  Since some post-AGB evolution codes tie the pulsation period to a mass loss prescription (i.e., Bl\"{o}cker) that assumes that intensive mass loss has ended by
the time that the pulsation period has reduced to 100 d, these models will need to be revised.
This would have the result of significantly increasing the initial stages of post-AGB evolution.

6. An average rate of change of the pulsation period with time can be determined by combining the empirically-determined rate of variation of period with temperature and the model-dependent rate of change of temperature with time.  These lead to an approximate rate of period change of $-$0.2 d/yr, a rate that is potentially detectable within a few decades.  Our observations of individual PPNs over this 14 year interval are complicated by the evidence for multiple periods, but give some tentative results that are not inconsistent with this value.

As can be seen, the study of the light curves of PPNs and the determination of their periods has the potential to constrain the models of post-AGB stellar evolution and to reveal evolution in real time.  In addition, these light curves together with radial velocity curves, when compared with appropriate pulsation models, have the potential to determine fundamental properties of the stars, such as mass and luminosity.  These will be explored further in subsequent papers.  
We are continuing to monitor the light curves of many of this PPNs to extend the time baseline.

\acknowledgments We begin by acknowledging the following additional VU undergraduate summer research students who participated in this long-term research program: 
Danielle Boyd, Laura Nickerson, Paul Barajas, Jason Webb, 
George Lessmann, Will Herron, Emily Cronin, Ryan Doering, Shannon Pankratz, Andrew Juell, Daniel Allen, Justin Lowry, Kathy Cooksey, Jeffrey Eaton, Nicolas George, Katie Musick, Sarah Schlobohm, Brian Bucher, Kara Klinke, Kristina Wehmeyer, Bradley Rush, Byung-Hoon Yoon, Jeffrey Massura, Marta Stoeckel, Larry Selvey, Jason Strains, Ansel Hillmer, Erin Lueck, Joseph Malan, and Callista Steele.
We thank Todd Hillwig and Detlef Sch\"{o}nberner for helpful comments on the manuscript
and Kevin Volk for ongoing conversations on a range of related topics.
We also thank the anonymous referee for his/her suggestions which improved the presentation
of these results.
The ongoing work of Paul Nord in maintaining the equipment is gratefully acknowledged.
Equipment support for the VU Observatory was partially provided by grants from
National Science Foundation College Science Instrumentation Program (8750722), the Lilly ``Dream of Distinction'' Program, and the Juenemann Foundation.  BJH acknowledges onging support from the National Science
Foundation (9315107, 9900846, 0407087), NASA through the JOVE program, and the Indiana
Space Grant Consortium.
This research has made use of the SIMBAD database, operated at CDS, Strasbourg,
France, and NASA's Astrophysical Data System.

\clearpage

\begin{deluxetable}{crrrrrcl}
\rotate
\tablecaption{List of C-Rich PPNs Observed\label{object_list}}
\tabletypesize{\footnotesize} \tablewidth{0pt} \tablehead{
\colhead{IRAS ID}&\colhead{GSC ID\tablenotemark{a}}&\colhead{2MASS ID}&\colhead{R.A.(2000.0)\tablenotemark{b}}&\colhead{Decl.(2000.0)\tablenotemark{b}}
&\colhead{V(mag)}&\colhead{Sp.T.}&\colhead{Other ID}} \startdata
Z02229+6208\tablenotemark{c} & 04050$-$02366 & J02264179+6221219 & 02:26:41.8 & +62:21:22 & 12.1 & G8-K0 0-Ia:& \nodata \\
04296+3429  & 02381$-$01014 & J04325697+3436123 &04:32:57.0 & +34:36:12 & 14.2 & G0 Ia & \nodata \\
05113+1347  & 00711$-$00365 & J05140775+1350282 &05:14:07.8 & +13:50:28 & 12.4 & G8 Ia & \nodata \\
05341+0852  & 00701$-$00027 & J05365506+0854087 &05:36:55.1 & +08:54:09 & 13.6 & G2 0-Ia: & \nodata \\
07134+1005  & 00766$-$00782 & J07161025+0959480 &07:16:10.3 & +09:59:48 & 8.2 & F5 I & HD~56126, CY CMi \\
07430+1115  & 00782$-$00087 & J07455139+1108196 &07:45:51.4 & +11:08:29 & 12.6 & G5 0-Ia: & \nodata \\
19500$-$1709 & 06317$-$00218 & J19525269$-$1701503 &19:52:52.7 & $-$17:01:50 & 8.7 & F3 I & HD~187885, V5122 Sgr \\
20000+3239  & 02674$-$02983 & J20015951+3247328  & 20:01:59.5 & +32:47:33 & 13.3 & G8 Ia & \nodata \\
AFGL~2688\tablenotemark{d}   & 02713$-$01972 &J21021878+3641412 &21:02:18.8 & +36:41:41 & 12.2 & F5~Iae & Egg~Nebula, V1610~Cyg \\
22223+4327  & 03212$-$00676 &J22243142+4343109 & 22:24:31.4 & +43:43:11 & 9.7 & G0 Ia & DO 41288, V448 Lac \\
22272+5435  & 03987$-$01344 &J22291039+5451062 & 22:29:10.4 & +54:51:06 & 9.0 & G5 Ia & HD~235858, V354~Lac \\
23304+6147  & 04284$-$00918 &J23324479+6203491 & 23:32:44.8 & +62:03:49 & 13.1 & G2 Ia & \nodata \\
\enddata
\tablenotetext{a}{Refers to the Hubble Space Telescope Guide Star Catalog.}
\tablenotetext{b}{Coordinates from the 2MASS Catalog.}
\tablenotetext{c}{Listed in the {\it IRAS} Faint Source Reject File (thus the ``Z'') but not in the Point Source Catalog. }
\tablenotetext{d}{Not included in the {\it IRAS} Point Source Catalog. }
\end{deluxetable}

\clearpage

\begin{deluxetable}{rlrccccc}
\tablecaption{List of Spectroscopic Properties of the C-Rich PPNs\tablenotemark{a}\label{object_prop} } \tablewidth{0pt}
\tablehead{ \colhead{IRAS
ID}&\colhead{Sp.T.}&\colhead{C/O}&\colhead{C$_2$}&\colhead{C$_3$}
&\colhead{AIBs}&\colhead{21 $\mu$m}&\colhead{30 $\mu$m}} \startdata
Z02229+6208 & G8-K0 0-Ia: & \nodata &Y& Y &Y &Y & Y \\
04296+3429  & G0 Ia & \nodata & Y& Y  & Y & Y & Y   \\
05113+1347  & G8 Ia & 2.4 & Y& Y &Y  & Y & Y  \\
05341+0852  & G2 0-Ia: &1.6 & Y& Y  &Y  & Y & Y   \\
07134+1005  & F5 I & 1.0 & Y& N &Y & Y & Y    \\
07430+1115  & G5 0-Ia: &\nodata & Y& Y &Y & Y & Y   \\
19500$-$1709 & F3 I & 1.0 & Y& N & Y\tablenotemark{b} & Y & Y   \\
20000+3239  & G8 Ia & \nodata & Y& \nodata &Y & Y & Y    \\
AFGL~2688 & F5~Iae & 1.0 & Y& Y &Y & Y & Y  \\
22223+4327  & G0 Ia & 1.2 & Y& Y & Y & Y & Y  \\
22272+5435  & G5 Ia & 1.6 & Y& Y & Y & Y & Y   \\
23304+6147  & G2 Ia & 2.8 & Y& Y & Y & Y & Y   \\
\enddata
\tablenotetext{a}{Taken from \citet{hri08a}.} 
\tablenotetext{b}{Weak plateau features.}
\end{deluxetable}

\clearpage

\begin{deluxetable}{lrrrrrcc}
\tablecaption{PPN Standard Magnitudes\tablenotemark{a}\label{std_ppn}}
\tabletypesize{\footnotesize} 
\tablewidth{0pt} \tablehead{ \colhead{IRAS ID} &\colhead{V}
&\colhead{B-V} &\colhead{V$-$R} &\colhead{R-I} &\colhead{V-I}
&\colhead{HJD$-$2,400,000} &\colhead{Ref.\tablenotemark{b}} } \startdata
Z02229$+$6208 & 12.09 & 2.83 & 1.68 & 1.49 & 3.17  & 1993 Oct 27  & 1 \\
04296$+$3429  & 14.21   & 1.99  & \nodata & \nodata & 2.57  & 1988 Oct 18   & 2  \\
05113$+$1347  & 12.41   & 2.13  & \nodata  & \nodata  & 2.15  & 1989 Oct 01  & 3  \\
05341$+$0852  & 13.55   & 1.81  & 1.09  & 1.03  & 2.12  & 1995 Sep 13   & 1  \\
07134$+$1005  & \nodata   & \nodata  & \nodata  & \nodata  & \nodata   & \nodata &   \\
07430$+$1115  & 12.62 & 1.86 & 0.99 & 0.94 & 1.93  & 1993 Oct 27  & 1\\
19500$-$1709  &  8.67 & 0.52  & \nodata & \nodata  & \nodata  &1987 Sep 04  & 4\\
			&  8.72 & \nodata  & 0.40 & \nodata  & \nodata  & 1995 Aug 22   & \\
20000$+$3239  & 13.17 & 2.77  & \nodata & \nodata  & 3.01  & 1988 Oct 18    & 3\\
			& 13.39 & 2.74  & 1.59 & 1.48  & 3.07  & 1989 Aug 24   & 3\\ 						& 13.26 & \nodata   & 1.65 & \nodata & \nodata  & 1995 Aug 22   & \\
              		& 13.27 & \nodata & 1.54 & \nodata & \nodata& 2002 Sep 12   &\\
AFGL~2688      & \nodata & \nodata & \nodata & \nodata & \nodata & \nodata  & \\
22223$+$4327  &  9.69 & 0.92 & 0.54 & 0.52  & 1.06 & 1989 Aug 24    & 3\\
			&  9.76 & \nodata & 0.54 & \nodata & 1.11 &2003 Jul 24  &\\
22272$+$5435  &  8.68 & 2.00  & \nodata & \nodata & 1.96  & 1988 Oct 18 & 5\\
			&  9.52 & 1.88  & 1.05 & 1.13  & 2.18  & 1989 Aug 24  & 5\\
			&  8.61 & \nodata & 1.01 &\nodata & \nodata & 1995 Oct 17  & \\
              		&  8.69 & \nodata & 1.03 & \nodata & 1.95 & 2003 Jul 24  &\\
23304$+$6147  & 13.06 & 2.31  & \nodata & \nodata & 2.63 & 1988 Oct 18  & 2 \\
 			& 13.15 & 2.37  & 1.36  & 1.29  & 2.65 & 1989 Aug 24 &  2 \\
              		& 12.99 & \nodata & 1.34  &\nodata & \nodata & 1995 Oct 17 &  \\
\enddata
\tablecomments{Uncertainties in the brightness and color are $\pm$0.01$-$0.02 mag.}
\tablenotetext{a}{Magnitudes of the PPNs on the standard BVRI system at the time of the observations. }
\tablenotetext{b}{References for our previously published photometry: (1) \citet{hri99}, (2) \citet{hri91a}, (3) \citet{kwo95}, (4) \citet{hri89}, (5) \citet{hri91b}. }
\end{deluxetable}

\clearpage

\begin{deluxetable}{llrrrrrrl}
\tablecaption{Comparison Star Identifications and Standard Magnitudes\label{std_comp}}
\tablewidth{0pt} \tablehead{ \colhead{IRAS Field}
&\colhead{Object} &\colhead{GSC ID} &\colhead{V} &\colhead{B-V}
&\colhead{V$-$R} &\colhead{R-I} &\colhead{V-I}
&\colhead{Observatory\tablenotemark{a}}} \startdata
Z02229$+$6208 & C1 & 04050-02770 & 10.85 & ...  & 0.40 & 0.30 & ...  & KPNO$-$1993 \\
              & C2 & 04050-02263 & 11.60 & ...  & 1.02 & 0.89 & ...  & KPNO$-$1993 \\
              & C3 & 04050-02497 & ...   & ...  & ...  & 0.30 & ...  & KPNO$-$1993 \\                           
04296$+$3429  & C1 & 02381-00100 & ...   & ...  & ...  & ...  & ...  & ... \\
	      & C2 & 02381-00277 & ...   & ...  & ...  & ...  & ...  & ... \\
              & C3\tablenotemark{b} & 02381-00374 & ...   & ...  & ...  & ...  & ...  & ... \\
05113$+$1347 & C1 & 00711-00461 & ...   & ...  & ...  & ...  & ...  & ... \\
              & C2 & 00711-00585 & ...   & ...  & ...  & ...  & ...  & ... \\
              & C3\tablenotemark{b} & 00711-00263 & ...   & ...  & ...  & ...  & ...  & ... \\
05341$+$0852  & C1 & 00701-00111 & 11.82 & ...  & ...  & ...  & 0.77 & KPNO$-$1995 \\
              & C2 & 00701-00867 & ...   & ...  & ...  & ...  & ...  & ... \\
              & C3\tablenotemark{b} & 00701-00579 & 12.07 & ...  & 0.33 & ...  & 0.63 & KPNO$-$1995 \\
07134$+$1005  & C1\tablenotemark{c} & 00766-02128 & ...   & ...  & ...  & ...  & ...  & ... \\
              & C2\tablenotemark{c} & 00766-00892 & ...   & ...  & ...  & ...  & ...  & ... \\
              & C3 & 00766-01866 & ...   & ...  & ...  & ...  & ...  & ... \\
07430$+$1115 & C1 & 00782-00165 & ...   & ...  & ...  & ...  & ...  & ... \\
              & C2 & 00782-00061 & ...   & ...  & ...  & ...  & ...  & ... \\
              & C3 & 00782-00007 & 11.97 & 0.45 & 0.28 & 0.25 & ...  & KPNO$-$1993 \\
19500$-$1709  & C1 & 06317-01314 & 10.08 & ...  & 0.26 & ...  & ...  & VUO$-$1995 \\
              & C2\tablenotemark{b} & 06317-00386 & 09.97 & ...  & 0.40 & ...  & ...  & VUO$-$1995 \\
              & C3 & 06317-00862 & 11.70 & ...  & 0.91 & ...  & ...  & VUO$-$1995 \\
20000$+$3239  & C1 & 02674-00275 & 11.43 & ...  & 0.60 & ...  & ...  & VUO$-$1995,2002 \\
              & C2 & 02674-00602 & 11.54 & ...  & 0.65 & ...  & ...  & VUO$-$1995,2002 \\
              & C3\tablenotemark{b} & 02674-01238 & 10.55 & ...  & 0.36 & ...  & ...  & VUO$-$1995,2002 \\
AFGL~2688      & C1 & 02713-02084 & 12.20 & ...  & 0.74 & ...  & ...  & VUO$-$2002 \\
              & C2 & 02713-01684 & 11.75 & ...  & 0.10 & ...  & ...  & VUO$-$2002 \\
              & C3\tablenotemark{b} & 02713-01106 & 12.55 & ...  & 0.34 & ...  & ...  & VUO$-$2002 \\
22223$+$4327  & C1 & 03212-00672 & 11.09 & ...  & 0.32 & ...  & 0.66 & VUO$-$2003 \\
              & C2 & 03212-00561 & 11.89 & ...  & 0.20 & ...  & 0.45 & VUO$-$2003 \\
              & C3 & 03212-00499 & 12.20 & ...  & 0.63 & ...  & 1.23 & VUO$-$2003 \\
22272$+$5435  & C1 & 03987-00512 & 11.17 & ...  & 0.59 & ...  & 1.21 & VUO$-$1995,2003 \\
              & C2 & 03987-02250 & 12.76 & ...  & 0.12 & ...  & 0.37 & VUO$-$1995,2003 \\
              & C3 & 03987-01072 & 12.69 & ...  & 0.16 & ...  & 0.46 & VUO$-$1995,2003 \\
23304$+$6147  & C1 & 04284-00700 & 12.73 & ...  & 0.46 & ...  & 0.86 & CFHT$-$1991,VUO$-$1995 \\
              & C2 & 04284-01228 & 12.73  & ...  & 0.45  & ...  & ...  & VUO$-$1995 \\
              & C3\tablenotemark{b}  & 04284-01162 & 12.44 & ...  & 0.34 & ...  & ...  & VUO$-$1995 \\
\enddata
\tablecomments{Uncertainties in the brightness and color are $\pm$0.01$-$0.02 mag.}
\tablenotetext{a}{Name of observatory and year of observation: VUO = Valparaiso University Observatory, 
KPNO = Kitt Peak National Observatory, CFHT = Canada-France-Hawaii Telescope. }
\tablenotetext{b}{Several of the comparison stars show low-amplitude, long-term variations over the 14 years of observations, with maximum peak-to-peak {\it V} amplitudes as follows: 
04296$-$C3 (0.04 mag), 05113$-$C3 (0.04 mag), 05341$-$C3 (0.015 mag), 
19500$-$C2 (0.030 mag)
20000$-$C3 (0.040 mag), AFGL 2688$-$C3 (0.015 mag), 23304$-$C3 (0.025 mag). }
\tablenotetext{c}{Suggestion of low-level, intrinsic variability in C1 or C2 within a season. }
\end{deluxetable}

\clearpage

\tablenum{5}
\begin{deluxetable}{rrrrrr}
\tablecaption{Differential Standard Magnitudes and Colors for
IRAS 22223+4327\tablenotemark{a} \label{tab_22223}}
\tablewidth{0pt} \tablehead{ \colhead{HJD$-$2,400,000}
&\colhead{$\Delta$V} 
&\colhead{HJD$-$2,400,000} &\colhead{$\Delta$R$_C$}
 &\colhead{HJD$-$2,400,000}
&\colhead{$\Delta$(V$-$R$_C$)} }
\startdata
49538.7901 & -1.415 & 49744.5037 & -1.578 & 49744.4992  & 0.239  \\
49551.8378 & -1.341 & 49874.8212 & -1.651 & 49874.8210  & 0.195  \\
49552.8491 & -1.343 & 49883.8492 & -1.611 & 49883.8490  & 0.200  \\
49558.8055 & -1.324 & 49929.8553 & -1.606 & 49929.8549  & 0.223  \\
49559.7618 & -1.325 & 49938.7664 & -1.636 & 49938.7661  & 0.214  \\
49563.7616 & -1.331 & 49945.7663 & -1.639 & 49945.7666  & 0.204  \\
49564.8120 & -1.326 & 49952.7807 & -1.646 & 49952.7809  & 0.190  \\
49565.7795 & -1.339 & 50007.6637 & -1.569 & 50007.6604  & 0.234  \\
49571.7768 & -1.344 & 50276.8069 & -1.602 & 50276.8069  & 0.194  \\
49580.7788 & -1.418 & 50289.8411 & -1.630 & 50289.8411  & 0.203  \\
\enddata
\tablenotetext{a}{Tables 5$-$17 are published in their entirety in the
electronic edition of the Astrophysical Journal.  A portion of Table~\ref{tab_22223} is
shown here for guidance regarding form and content.}
\end{deluxetable}

\clearpage

\tablenum{18}
\begin{deluxetable}{rrrrrrrrrrrrr}
\tablecolumns{13} \tabletypesize{\scriptsize}
\tablecaption{Statistics of the PPN Light Curves\label{tab_stat}}
\tabletypesize{\footnotesize} 
\tablewidth{0pt} \tablehead{ 
\colhead{} && \multicolumn{3}{c}{Number of Observations} && 
\multicolumn{3}{c}{Maximum Uncertainty\tablenotemark{a}} && \multicolumn{3}{c}{Average Uncertainty\tablenotemark{a}} \\
\cline{3-5} \cline{7-9} \cline{11-13}
\colhead{IRAS ID} &&
\colhead{$\Delta$V}&\colhead{$\Delta$R} &\colhead{$\Delta$(V$-$R)}&&
\colhead{$\sigma$($\Delta$V)} &\colhead{$\sigma$($\Delta$R)} &\colhead{$\sigma$($\Delta$(V-R))}
&&\colhead{$\sigma$($\Delta$V)} &\colhead{$\sigma$($\Delta$R)} &\colhead{$\sigma$($\Delta$(V-R))}} \startdata
Z02229$+$6208 && 130 & 88 & 62&& 0.013 & 0.010  & 0.014  && 0.005 & 0.003 & 0.006\\
04296$+$3429  && 70   & 39  & 28  && 0.030 & 0.016 & 0.028  && 0.020 & 0.010 & 0.020\\
05113$+$1347  && 78   & 37  & 31 && 0.014  & 0.009 & 0.017  && 0.008  & 0.005  & 0.009\\
05341$+$0852  && 67   & 35  & 31  && 0.025  & 0.011 & 0.026 && 0.014  & 0.008  & 0.017\\
07134$+$1005  && 89   & 58   & 48  && 0.010 & 0.008 & 0.008 && 0.004  & 0.004  & 0.005\\
07430$+$1115  &&  73 & 47  & 39 && 0.010 & 0.008  & 0.010 && 0.006 & 0.003  & 0.006\\
19500$-$1709  && 181  & 125 & 98 && 0.007 & 0.007 & 0.008 && 0.003 & 0.003 & 0.003\\
20000$+$3239  && 213 & 173  & 131 && 0.017 &0.012  & 0.020 && 0.010 &0.005  & 0.011\\
AFGL~2688-N  && 135 & 101  & 75  && 0.011 & 0.010 & 0.015 && 0.006 & 0.005  & 0.007\\
AFGL~2688-S  &&  124   & 99  & 69  && 0.020 & 0.018 & 0.024 && 0.011 & 0.008 & 0.012\\
22223$+$4327  && 267 & 212  & 160 && 0.008  & 0.008  & 0.011 && 0.003 & 0.003 &0.004\\
22272$+$5435  && 248 & 189  & 132 && 0.008  & 0.008 & 0.011 && 0.004 & 0.004 & 0.006\\
23304$+$6147  &&197 &163  & 111  && 0.020 & 0.020 & 0.023 && 0.011 & 0.009 & 0.013\\
\enddata
\tablenotetext{a}{The maximum or average uncertainty in a single differential measurement.}
\end{deluxetable}

\clearpage

\tablenum{19}
\begin{deluxetable}{rrrrcrrrrrrr}
\tablecaption{Results of the Period Study of Carbon-Rich PPNs\label{P_results}}
\rotate
\tabletypesize{\footnotesize} 
\tablewidth{0pt} \tablehead{ 
\colhead{} & \multicolumn{3}{c}{CLEAN Analysis} && 
\multicolumn{7}{c}{Period04 Analysis}   \\
\cline{2-4} \cline{6-12} 
\colhead{} & \colhead{V}&\colhead{R} &\colhead{V$-$R} 
&&\colhead{V} &\colhead{V} &\colhead{V}
&\colhead{R}&\colhead{R}&\colhead{V$-$R}&\colhead{V$-$R}  \\
\colhead{} & \colhead{(1994$-$2008)}&\colhead{(1994$-$2008)} &\colhead{(1994$-$2008)} 
&&\colhead{(1994$-$2008)} &\colhead{(1994$-$1999)} &\colhead{(2002.6$-$2008)}
&\colhead{(1994$-$2008)}&\colhead{(2002.6$-$2008)}&\colhead{(1994$-$2008)}&\colhead{(2002.6$-$2008)} \\
 \colhead{IRAS ID} &\colhead{P(day)} &\colhead{P(day)} &\colhead{P(day)} & &\colhead{P(day)} &\colhead{P(day)} &\colhead{P(day)} &\colhead{P(day)} & 
 \colhead{P(day)} &\colhead{P(day)} &\colhead{P(day)}  
 } \startdata
Z02229$+$6208 & 153.0,137.1,813 & 153.3,137.6 & 152.8 &&153.0,137.1$\pm$0.4,854 & 152.4$\pm$1.2,820 & 136.2,152.3$\pm$0.9  &153.3,137.0$\pm$0.6 & 136.1$\pm$1.0 & 152.6$\pm$0.5 & \nodata \\
04296$+$3429\tablenotemark{a}  & 71:   & \nodata  & \nodata  && 71:& \nodata & \nodata & \nodata & \nodata &  \nodata  &\nodata\\
05113$+$1347  & 133.1 & 137.8,127.8 & \nodata && 133.0$\pm$0.3  & \nodata &136.4$\pm$0.9 & 137.4$\pm$0.4 & 136.2$\pm$0.9 &\nodata&\nodata \\
05341$+$0852\tablenotemark{a}  & 93.8   & 92.8\tablenotemark{b}  & \nodata  && 93.9$\pm$0.1  & \nodata  & \nodata & \nodata & 92.9$\pm$0.8\tablenotemark{b} & \nodata &\nodata \\
07134$+$1005  & 35.1,44.4   & 38.8,49.8   & 35.0,528  && 35.1,44.5$\pm$0.1\tablenotemark{c}  & 44.1.38.9$\pm$0.1\tablenotemark{c} & 35.1$\pm$0.1\tablenotemark{c} & 38.8$\pm$0.1\tablenotemark{c} & 38.7.35.0$\pm$0.1\tablenotemark{c} &35.0$\pm$0.1\tablenotemark{c}&\nodata \\
07430$+$1115\tablenotemark{a}  &  135.9,148.5 & 98.9,136.2 & \nodata && 135.8,148.5$\pm$0.3 & \nodata  & \nodata & \nodata &\nodata & \nodata &\nodata\\
19500$-$1709\tablenotemark{a} & 38.3  & 38.4 &38.2,30.3 && 38.3,36.8,45.4$\pm$0.1\tablenotemark{c} & 50.7,36.8$\pm$0.1\tablenotemark{c}  & 42.4,38.2$\pm$0.1\tablenotemark{c} & 38.4$\pm$0.1\tablenotemark{c} & 38.2$\pm$0.1\tablenotemark{c} &\nodata & 7.4,30.5,38.0$\pm$0.1\tablenotemark{c}\\
20000$+$3239  & 152.6 & 152.6  & 155.0 && 152.7$\pm$0.3  & \nodata  & 153.5$\pm$0.8 & 153.0$\pm$0.5 & 153.3$\pm$0.9  & 155.0$\pm$0.4 &155.4$\pm$1.2\\
AFGL~2688-N\tablenotemark{a}  &  93.0,602 &  89.1,595  &  \nodata  && 93.2,89.5$\pm$0.2,600  & \nodata   & \nodata  & 61.7,89.1$\pm$0.2,595 & \nodata &  \nodata & \nodata \\
22223$+$4327\tablenotemark{a}  & 88.0,90.8 & 91.0  & 88.1,91.0 && 87.8,90.9$\pm$0.1  & 88.8$\pm$0.1  & 91.5,86.5$\pm$0.4 & 90.7$\pm$0.1 & 91.4$\pm$0.4 & 87.8,90.9$\pm$0.2 &89.2$\pm$0.2\\
22272$+$5435  & 132.2,124.9/127.7 & 127.2  & 127.5 && 132.0$\pm$0.1 & 133.1$\pm$0.4 & 128.1$\pm$0.3 & 127.7$\pm$0.3 & 128.0$\pm$0.3 & 127.3$\pm$0.2 &127.6$\pm$0.3\\
23304$+$6147  & 84.6 & 85.0  & \nodata  &&84.6$\pm$0.1  & 84.2$\pm$0.2    &66.8$\pm$0.2 & 85.0$\pm$0.1 & 81.5$\pm$0.3 & \nodata &\nodata\\
\enddata
\tablenotetext{a}{Light curve trend removed before analysis. }
\tablenotetext{b}{For the interval 2002$-$2006.}
\tablenotetext{c}{The uncertainties appear to be unrealistically small and are rounded up to $\pm$0.1 d.}
\end{deluxetable}

\clearpage

\tablenum{20}
\begin{deluxetable}{rrlllcll}
\tablecaption{Results of the Period and Light Curve Study of Carbon-Rich PPNs\tablenotemark{a}\label{results}}
\rotate
\tabletypesize{\footnotesize}
\tablewidth{0pt} \tablehead{ \colhead{IRAS ID} &\colhead{P}
&\colhead{SpT\tablenotemark{b}} &\colhead{T$_{\rm eff}$\tablenotemark{c}} 
&\colhead{$\Delta$V\tablenotemark{d}}&\colhead{$\Delta$(V$-$R)\tablenotemark{d}}
&\colhead{$\Delta$T$_{\rm eff}$\tablenotemark{d}}&\colhead{Comments}\\
 &\colhead{(day)}  
 & &\colhead{(K)} &\colhead{(mag)} &\colhead{(mag)} &\colhead{(K)} &  } \startdata
Z02229$+$6208 & 153 & G8-K0 Ia & 5500  & 0.54 & 0.06 & 400 & \nodata \\
05113$+$1347  & 133  & G8 Ia  & 5250  & 0.67 & 0.18 & 670 &\nodata \\
20000$+$3239  & 153 & G8 Ia  & 5250$-$5500\tablenotemark{e}   & 0.59 & 0.10 & 400 & \nodata \\
22272$+$5435  & 130 & G5 Ia  & 5750     & 0.49  & 0.12 & 750 & \nodata \\
07430$+$1115  &  136:\tablenotemark{f}& G5 0-Ia & 6000  & 0.23 & 0.05 & 310 & \nodata \\
05341$+$0852  & 94    & G2 Ia  & 6500  & 0.13 & 0.06 & 440 & brightening since 2005\\
23304$+$6147  & 85  &G2 Ia  & 6750    & 0.20 & 0.10 & 730 & \nodata \\
22223$+$4327  & 89 & G0 Ia  & 6500  & 0.21 & 0.06 & 440 & dimming trend \\
04296$+$3429  & 71:   & G0 Ia\tablenotemark{g} & 7000 & 0.13: & 0.07 & 550 &   brightening trend \\
AFGL~2688-N  &  $\sim$91:\tablenotemark{h}  & F5 Iae  & 6500  & 0.18 & 0.05 & 370 & brightening trend \\
AFGL~2688-S  &  \nodata   & F5 Iae  & 6500  & 0.30: & 0.07: &  \nodata & brightening trend \\
07134$+$1005  & $\sim$39  & F5 I  & 7250  & 0.18 & 0.04 & 340 &\nodata  \\
19500$-$1709  & 38   & F3 I & 8000  & 0.13 & 0.04 & 510 & sudden dimming, then brightening \\
\enddata
\tablenotetext{a}{Listed in order of spectral type.}
\tablenotetext{b}{Spectral types based on low-resolution spectra, primarily from \citet{hri95}.}
\tablenotetext{c}{T$_{\rm eff}$ determined from model atmosphere analyses by \citet{vanwin00}, \citet{red02}, \citet{red99}, and \citet{klo00}}
\tablenotetext{d}{The maximum brightness and color range observed in a season and the corresponding maximum temperature change.}
\tablenotetext{e}{No model atmosphere analysis; T$_{\rm eff}$ assumed to be similar to other G8 Ia PPNs in this sample.}
\tablenotetext{f}{An almost equally probably period is 148 d.}
\tablenotetext{g}{F3~I by \citet{sancon08}.}
\tablenotetext{h}{A period of $\sim$600 d also appears to be present in the data.}
\end{deluxetable}

\clearpage

\clearpage

\clearpage

\clearpage

\clearpage

\clearpage

\clearpage

\clearpage

\clearpage

\clearpage

\clearpage

\clearpage

\clearpage

\clearpage

\clearpage

\clearpage

\clearpage

\clearpage

\clearpage

\end{document}